\documentclass[showpacs
]{revtex4}
\usepackage{amssymb}
\usepackage{graphicx}
\usepackage{dcolumn}
\usepackage{bm}
\usepackage{epsfig}
\usepackage{amssymb}
\usepackage{color}

\newcommand{\bal}{\begin{align}}
\newcommand{\eal}{\end{align}}
\newcommand{\beq}{\begin{eqnarray}}
\newcommand{\eeq}{\end{eqnarray}}
\newcommand{\nneeq}{\nonumber \end{eqnarray}}

\newcommand{\nn}{\nonumber \\}
\newcommand{\es}{& = &}
\newcommand{\rs}{\, = \,}
\newcommand{\ps}{& + &}
\newcommand{\ms}{& - &}
\newcommand{\ts}{& \times &}
\newcommand{\nt}{\nn \ts}
\newcommand{\np}{\nn \ps}
\newcommand{\nm}{\nn \ms}

\newcommand{\cA}{ {\cal A} }
\newcommand{\cB}{ {\cal B} }

\newcommand{\cM}{ {\cal M} }
\newcommand{\cH}{ {\cal H} }
\newcommand{\cE}{ {\cal E} }

\newcommand{\cG}{ {\cal G} }
\newcommand{\cT}{ {\cal T} }

\newcommand{\cU}{ {\cal U} }

\newcommand{\cY}{ {\cal Y} }

\begin{document}
\title{ Asymptotic freedom in the front-form Hamiltonian \\
for quantum chromodynamics of gluons  }
\author{       Mar\'ia G\'omez-Rocha }
\email{maria.gomez-rocha@uni-graz.at}
\affiliation{  Institute of High Energy Physics, 
               Austrian Academy of Science,
               Nikolsdorfergasse 18, 1050 Vienna, Austria }
               \affiliation{Institute of Physics, University of Graz, NAWI Graz, A-8010 Graz, Austria}
               
\author{       Stanis{\l}aw D. G{\l}azek }
\email{stglazek@fuw.edu.pl}
\affiliation{  Institute of Theoretical Physics,
               Faculty of Physics, 
               University of Warsaw,
               Pasteura 5, 02-093 Warsaw, Poland }   
\date{        \today                 }
\begin{abstract}
Asymptotic freedom of gluons in QCD is obtained in 
the leading terms
of their renormalized Hamiltonian in the Fock space,
instead of considering virtual 
Green's functions or scattering amplitudes. Namely, 
we calculate the three-gluon interaction term in the
front-form Hamiltonian for effective gluons in the
Minkowski space-time using the renormalization group 
procedure for effective particles (RGPEP), with a new 
generator. The resulting three-gluon vertex is a function 
of the scale parameter, $s$, that has an interpretation 
of the size of effective gluons. The corresponding 
Hamiltonian running coupling constant, $g_\lambda$, 
depending on the associated momentum scale $\lambda 
= 1/s$, is calculated in the series expansion in powers 
of $g_0 = g_{\lambda_0}$ up to the terms of third
order, assuming some small value for $g_0$ at some 
large $\lambda_0$. The result exhibits the same 
finite sensitivity to small-$x$ regularization as the 
one obtained in an earlier RGPEP calculation, but the 
new calculation is simpler than the earlier one because 
of a simpler generator. This result establishes a degree 
of universality for pure-gauge QCD in the RGPEP. 
\end{abstract} 
\maketitle

\section{Introduction}
\label{intro}

This article describes a calculation of asymptotic
freedom~\cite{Gross:1973id, Politzer:1973fx,
PolitzerScattering, GrossWilczekPRDII} in the
leading ultraviolet terms of front form~\cite{Dirac1} 
(FF) Hamiltonians for gluons, using the renormalization
group procedure for effective particles (RGPEP, see
Sec.\ref{RGPEP}) developed in recent years as an
element of the program of constructing
non-perturbative QCD outlined in
Ref.~\cite{Wilsonetal}. Besides the asymptotically
free ultraviolet behavior, the calculation also
confirms finite dependence of the effective
Hamiltonian three-gluon coupling constant on the
regularization of small-$x$ singularities in the
bare theory. This dependence is of interest since the
small-$x$ behavior may in general be thought related 
to the vacuum state in the instant form~\cite{Dirac1}
(IF) of dynamics and in that form the vacuum is 
believed to be responsible for symmetry breaking and 
confinement. But the third-order RGPEP calculation 
reported here would have to be extended to higher 
orders to verify if it can shed any light on the 
relevant mechanisms.

In the earlier RGPEP
calculation~\cite{Glazek:2000dc}, a generator is
used that is suitable for perturbative
calculations but difficult to use beyond the
perturbative regime. In this article, we use a
generator that is much easier to use beyond
perturbative expansion. The difference between the
generators is further explained in
Sec.~\ref{RGPEP}. At the same time, our
perturbative calculation demonstrates that the new
RGPEP generator passes the test of producing
asymptotic freedom, which any method aiming at
solving QCD must pass. In particular, passing this
test is a precondition for tackling
non-perturbative issues, such as the ones that
emerge when one allows effective gluons to have
masses~\cite{Wilsonetal}.

An additional point is thus made that two
different versions of the RGPEP, defined using two
different generators, yield the same behavior of
the running coupling constant in three-gluon
Hamiltonian interaction term when the calculation
is carried out in the third-order expansion in
powers of the coupling constant. This means that a
considerable change in the the RGPEP generator
does not influence the finite behavior of the
three-gluon term. Therefore, we suggest that there
exists certain degree of universality in the
behavior of FF Hamiltonians in the RGPEP: the
leading terms in the Hamiltonian beta function are
universal and they are universally obtained using
different versions of the RGPEP.

The paper is organized as follows. The RGPEP is
generally but briefly described in
Sec.~\ref{RGPEP}. The bare Hamiltonian for gluons
is introduced in Sec.~\ref{BareHamiltonian}, which
includes the derivation of its FF density from the
Lagrangian density, solving constraints,
quantization, and regularization.
Sec.~\ref{ThreeGluonTerm} explains our third-order
calculation of the effective Hamiltonian
three-gluon interaction term as a function of the
RGPEP scale. This scale is denoted by $t =
\lambda^{-4}$, where $\lambda=1/s$ is the
invariant mass width of the form factors in
effective Hamiltonian vertices that solve the
RGPEP equation, and $s$ is the parameter that has
the interpretation of size of effective gluons.
The calculation includes identification of
second-order mass counterterms in
Sec.~\ref{massct}, third-order counterterm for the
three-gluon term in Sec.~\ref{gvct}, and the
definition and result one obtains for the
Hamiltonian running coupling constant in
Sec.~\ref{runningcoupling}. Comparison of the
previous and current calculation is summarized in
Sec.~\ref{Universality}, and Sec.~\ref{Conclusion}
concludes the paper. Several appendices provide
details necessary for completeness of the paper,
including details of the RGPEP in
App.~\ref{AppRGPEP}, details of the initial
Hamiltonian in App.~\ref{initialH}, integration of
the RGPEP equation order-by-order in
App.~\ref{RGPEPsolution}, calculation of the
three-gluon vertex counterterm in
App.~\ref{threegCT}, and method of evaluating the
third-order contributions to the running coupling
constant $g_\lambda$ with a formula explaining the
infrared stability guaranteed by the design of the
RGPEP in App.~\ref{Limitk12}.

\section{ The method of calculation }
\label{RGPEP}

In distinction from calculations of Euclidean
Green's functions~\cite{Gross:1973id,
Politzer:1973fx, PolitzerScattering,
GrossWilczekPRDII} and from early calculations
using infinite momentum and light-front
techniques~\cite{Casher, Thorn2, BrodskyLepage,
LepageBrodsky, Perry1}, or other approaches,
e.g.~\cite{Drell:1981gu, PMC, Alexanian, Szczepaniak,
Reinhardt}, some discussing three-gluon 
coupling~\cite{Eichmann, Blum:2014gna}, asymptotic freedom of gluons
is derived here as a feature of the Minkowski
space-time FF Hamiltonian that acts in the Fock
space of virtual gluons obtained as a result of an
explicit operator renormalization group
transformation. The transformation does not
involve any wave function renormalization constant
and the counterterms are calculated without
assuming multiplicative renormalizability. More
specifically, we use the RGPEP mentioned in
Sec.~\ref{intro} and we calculate the coefficient
in front of the Hamiltonian operator interaction
term that annihilates one effective gluon and
creates two, or annihilates two and creates one.
This coefficient is called the coupling constant.
It is denoted by $g_t$, since it depends on the
RGPEP scale parameter $t$ that corresponds to the
size $s$ of effective gluons, $t = s^4$. In the
calculated effective Hamiltonian operator, the
gluon interaction vertices are softened by the
form factors of width $\lambda = 1/s$ in momentum
variables. Asymptotic freedom is exhibited in the
behavior of the coupling constant $g_t$ as $t$
approaches zero, or $\lambda$ tends to infinity.

The size parameter for gluons is introduced by 
solving the RGPEP differential equation, 
\beq
\label{H'text}
{d \over dt} \cH_t 
\es \left[ \cG_t ,
\cH_t \right] \ ,
\eeq
where $\cH_t$ denotes the Hamiltonian of interest and
$\cG_t$ plays the role of a generator of the required
transformation, see Appendix~\ref{AppRGPEP}. The 
transformation changes the bare creation and annihilation 
operators for point-like gluons of canonical QCD, which are
denoted by $a_0$ in reference to $t=0$, to the operators 
for effective gluons of finite size $s = t^{1/4}$ that are 
denoted by $a_t$,
\beq
\label{at}
a_t \es \cU_t \, a_0 \, \cU_t^\dagger  \ ,
\eeq
where
\beq
\label{Usolution}
\cU_t 
\es 
T \exp{ \left( - \int_0^t d\tau \, \cG_\tau
\right) } \ ,
\eeq
and $T$ denotes ordering in $\tau$. The initial
condition for solving Eq.~(\ref{H'text}) is
provided at $t=0$ by the canonical FF Hamiltonian
with modifications implied by its divergent nature. 
These modifications include the regularization 
factors in interaction vertices and the counterterms 
whose structure is found using solutions to Eq.~(\ref{H'text}). 

Our choice of the generator has the form of a 
commutator, similar to but different from
Wegner's~\cite{Wegner},
\beq
\label{Gt}
\cG_t \es [ \cH_f, \cH_{Pt} ] \ ,
\eeq
where operators $\cH_f$ and $\cH_{Pt}$ are defined 
using parts of the Hamiltonian $\cH_t$. The operator
$\cH_f$ is the free part of $\cH_t$ and $\cH_{Pt}$ 
is defined in terms of the interaction terms as 
explained in Appendix~\ref{AppRGPEP}. The definition 
of $\cH_{Pt}$ secures that $\cU_t$ is invariant with 
respect to the seven-parameter Poincar\'e subgroup  
that forms the kinematical symmetry group of the 
FF of Hamiltonian dynamics. The generator $\cG_t$ 
in Eq.~(\ref{Gt}) is simpler and more suitable for 
non-perturbative calculations than the one used in
Ref.~\cite{Glazek:2000dc} (see below).

Once the generator is chosen and the counterterms
that complete the definition of the initial
condition for $\cH_t$ at $t=0$ are found, the
Hamiltonian for effective gluons of size $s$ is
uniquely determined, up to the value of the
coupling constant $g_0$ at some arbitrarily
chosen value of $t=t_0$. Thus, the renormalized FF
Hamiltonian for QCD could in principle be defined
using the RGPEP without any reference to
perturbation theory. 

However, our calculation is only carried out using
expansion in powers of $g_t$~\cite{Glazek:2012qj}
up to the third order. The reason is that too
little is currently known about the counterterms
required for non-perturbative calculations. Even
the terms of second and third order require
study. We show in the next sections that the
generator $\cG_t$ of Eq.~(\ref{Gt}) produces the
same third-order dependence of $g_t$ on $t$ as the
one obtained in Ref.~\cite{Glazek:2000dc}. We thus
obtain the Minkowski space-time Hamiltonian
example of the universality of leading perturbative 
terms in the coupling constant in asymptotically 
free theories. Our perturbative calculation indicates
what kind of terms are necessary to counter the
ultraviolet divergences. They also show how the 
small-$x$ singularities appear in addition to the
ultraviolet ones, and that they cancel out, leaving 
behind finite effects. 

The current level of knowledge about the FF
Hamiltonians for QCD being quite limited,
calculations of higher order than the third one
discussed here are required to gain more
information. However, the higher-order
calculations require the third-order output
reported here as an input. Given the RGPEP
Eq.~(\ref{H'text}) and its systematic
expansion~\cite{Glazek:2012qj}, one may hope that
it will eventually become possible to identify the
structure of $\cH_t$ required for obtaining
non-perturbative solutions with mathematically
estimable precision, perhaps using conceptual
analogies between the RGPEP and procedures
discussed in Refs.~\cite{Wilson:1965zzb,
Wilson:1970tp}.

\section{ Canonical Hamiltonian for gluons }
\label{BareHamiltonian}

The initial condition for solving
Eq.~(\ref{H'text}) is the canonical FF Hamiltonian
for gluons in QCD plus counterterms. In this
section we describe the canonical Hamiltonian for
gluons. The description introduces the notation
for details that appear throughout the article.

The canonical Hamiltonian is derived from the standard 
Lagrangian density
\beq
 {\cal L} = - {1 \over 2} \text{tr} \, F^{\mu \nu}
F_{\mu \nu} \ ,
\eeq
where 
$F^{\mu \nu} = \partial^\mu A^\nu 
- \partial^\nu A^\mu + i g [A^\mu, A^\nu]$,  
$A^\mu = A^{a \mu} t^a$, 
$ [t^a,t^b] = i f^{abc} t^c$
and 
$\text{tr} \, t^a t^b = \delta^{ab}/2$.
The associated energy-momentum density tensor 
reads, 
\beq
\cT^{\mu \nu} \es
-F^{a \mu \alpha} \partial^\nu A^a_\alpha + g^{\mu \nu} F^{a \alpha
\beta} F^a_{\alpha \beta}/4 \ .
\eeq
The FF Hamiltonian is obtained by integrating 
the component $\cT^{+ \, -}$ over the hyperplane 
defined by the condition $ x^+ = x^0 + x^3 = 0$. 
We work in the gauge $A^+ = 0$, in which the 
Lagrange equations constrain $A^-$ to 
\beq
\label{A-1}
A^- \es 
{ 1 \over \partial^+ } \, 2 \, \partial^\perp A^\perp 
- { 2 \over \partial^{ + \, 2} } \ 
ig \, [ \partial^+ A^\perp, A^\perp] \ ,
\eeq
so that the only degrees of freedom are the fields
$A^\perp$. The first term in $A^-$ is independent
of the coupling constant $g$. This term is by
definition included in a new constrained field,
which is denoted by the same symbol $A$ in what
follows, with
\beq
\label{A-}
A^- \es { 1 \over \partial^+ } \, 2 \,
\partial^\perp A^\perp \ ,
\eeq
while the second term is explicitly included in
the interaction Hamiltonian that is written in
terms of fields $A^\perp$, using $A^-$ defined
in Eq.~(\ref{A-}). Employing this convention 
and freely integrating by parts, one obtains 
the FF energy of the constrained gluon field 
in the form
\begin{equation}
P^- = {1 \over 2}\int dx^- d^2 x^\perp {\cal H}\, |_{x^+=0} \quad ,
\end{equation}
where ${\cal H} = \cT^{+ -}$ is a sum of four terms,
denoted as in Ref.~\cite{Glazek:2000dc},
\beq
\cT^{+ -} = {\cal H}_{A^2} + {\cal H}_{A^3} + {\cal H}_{A^4} + {\cal
H}_{[\partial A A]^2} \ .
\eeq
The terms are~\cite{Casher,Thorn2,BrodskyLepage}
\beq
\label{HA2}
{\cal H}_{A^2} \es - {1\over 2} A^{\perp a } (\partial^\perp)^2 A^{\perp a} \  , \\
\label{HA3}
{\cal H}_{A^3} \es  g \, i\partial_\alpha A_\beta^a [A^\alpha,A^\beta]^a  \ , \\
\label{HA4}
{\cal H}_{A^4} \es  - {1\over 4} g^2 \, [A_\alpha,A_\beta]^a[A^\alpha,A^\beta]^a  \ , \\
\label{HA2A2}
{\cal H}_{[\partial A A]^2} \es  {1\over 2}g^2 \,
[i\partial^+A^\perp,A^\perp]^a {1 \over (i\partial^+)^2 }
[i\partial^+A^\perp,A^\perp]^a \ .
\eeq
The bare expression for the quantum gluon 
energy operator is obtained through replacing
$A^\mu$ in $\cT^{+ \, -}$ by an operator 
four-vector $\hat A^\mu$, which is defined 
by its Fourier composition on the front 
corresponding to $x^+=0$,
\beq
\hat A^\mu \es \sum_{\sigma c} \int [k] \left[ t^c \varepsilon^\mu_{k\sigma}
a_{k\sigma c} e^{-ikx} + t^c \varepsilon^{\mu *}_{k\sigma}
a^\dagger_{k\sigma c} e^{ikx}\right]_{x^+=0} \ .
\eeq
This operator acts in the Fock space spanned by states
created by products of the creation operators 
$a^\dagger_{k\sigma c}$ on the bare vacuum state 
$|0\rangle$. 

In the operator $\hat A^\mu$, the Fourier-like
integral over kinematical momentum variables is
carried out with the measure $[k] = \theta(k^+)
dk^+ d^2 k^\perp/(16\pi^3 k^+) $. Thus, the
integration matches the one in the Fourier
transform only in the transverse directions,
the integral over $k^+$ being limited to
positive values. 

The polarization four-vectors
$\varepsilon$ have components 
\beq
\varepsilon^\mu_{k\sigma} 
\es (\varepsilon^+_{k\sigma}=0, \varepsilon^-_{k\sigma} 
= 2k^\perp \varepsilon^\perp_\sigma/k^+, 
\varepsilon^\perp_\sigma) \ .
\eeq 
The symbol $\sigma$ labels the gluon spin polarization 
and $c$ is a color index. The creation and annihilation 
operators satisfy commutation relations
\beq
\left[ a_{k\sigma c}, a^\dagger_{k'\sigma' c'} \right] 
\es 
k^+
\tilde \delta(k - k') \,\, \delta^{\sigma \sigma'}
\, \delta^{c c'} \ ,
\eeq
where $\tilde \delta(p) = 16 \pi^3 \delta(p^+) \delta(p^1)
\delta(p^2)$, and commutators among all $a$s, and among all
$a^\dagger$s, vanish. By definition, $ a_{k\sigma c} 
|0\rangle = 0$ for all momenta, spins and colors. 

Normal-ordering of the operator density $\cH(\hat A)$ 
defines the FF integrand in the Hamiltonian, 
\beq
\label{hatP-}
\hat P^- \es {1 \over 2} \int dx^- d^2 x^\perp  \
: \cH(\hat A ) : \ ,
\eeq
in which all annihilation operators are on 
the right side of all creation operators. 
Details of $\hat P^-$ are given in 
Appendix~\ref{initialH}.

\subsection{ Regularization }
\label{reg}

The bare Hamiltonian is regularized by introducing
regulating factors, denoted by $r$, in the
interaction terms. These factors make the
interaction terms vanish~\cite{Glazek:2000dc} if
the associated change of any gluon relative
transverse momentum were to exceed the very large
cutoff parameter $\Delta$. Likewise, the
interactions are also made to vanish if any change
of any longitudinal momentum fraction $x$ of any
gluon involved in the interaction were to be
smaller than the very small cutoff parameter
$\delta$. Various regularization factors can be
incorporated in the interaction terms in $\hat
P^-$ of Eq.~(\ref{hatP-}) to realize these
conditions. Appendix~\ref{initialH} shows how the
regularization factors are introduced in $\hat
P^-$ according to the following rules.

In every interaction Hamiltonian term every
particle creation and annihilation operator is
labeled by its momentum quantum numbers $p^+$ and
$p^\perp$. Let the total momentum of all quanta
annihilated in a term have components $P^+$ and
$P^\perp$. These are the same as components of the
total momentum of quanta created in the term.
The relative momentum fraction $x$ for the quantum
of momentum $p$ is defined as the ratio
\beq
x_{p/P} \es p^+/P^+ \ ,
\eeq 
and the relative transverse momentum for the quantum 
is defined by 
\beq
\kappa_{p/P} \es p^\perp - x P^\perp \ .
\eeq
Every creation and annihilation operator in every 
term in the entire canonical Hamiltonian of any 
momentum $p$ is multiplied by the regulating factor 
\beq
r_{\Delta \delta}(\kappa^\perp, x) 
\es 
r_\Delta( \kappa^\perp ) r_\delta(x) \theta(x) \ .
\eeq
We use one transverse regulator factor 
\beq
r_\Delta(z) \es \exp{(- z/\Delta^2)} \ ,
\eeq
and one of the following three different small-$x$ 
regulator factors,
\beq
\label{r1}
\text a ) & & \quad  r_\delta (x) \rs x/(x+\delta) \ , \\
\label{r2}
\text b ) & & \quad  r_\delta (x) \rs \theta(x-\delta) \ , \\
\label{r3}
\text c ) & & \quad  r_\delta (x) \rs x^\delta \, \theta(x-\epsilon) \ .
\eeq
Dependence on the transverse regulator factors
will be removed using the RGPEP. Effects of
the small-$x$ regularization will be described 
by comparing results obtained using different 
regulator factors in Eqs.~(\ref{r1}) to
(\ref{r3}).

Regularization factors in canonical QCD terms
that are quartic in gluon field operators are 
additionally specified by treating every such 
term as built solely from vertices in which 
one quantum is changed to two or vice versa. 
This regularization choice also applies to the 
seagull terms that result from constraints on 
$A^-$, as if the constrained field component 
corresponded to an exchange of a quantum with 
a corresponding momentum. Details are available
in Appendix~\ref{initialH}. In an abbreviated 
notation, the regularization uses symbols
$\tilde r_{P,p} = \tilde r_{\Delta\delta}(P,p)$,
where 
\beq
\tilde r_{\Delta\delta}(P,p) 
\es r_{\Delta\delta}(p^\perp - x_{p/P} P^\perp, x_{p/P})
\,  r_{\Delta\delta}[P^\perp - p^\perp  -
(1-x_{p/P}) P^\perp, 1-x_{p/P} ] \ .
\eeq

\subsection{ Counterterms }
\label{cterms}

The initial condition for solving the RGPEP
Eq.~(\ref{H'text}) is provided by the regulated
canonical Hamiltonian plus counterterms. The
ultra-violet divergent parts of these
counterterms, depending on the regularization
parameter $\Delta$, are found in a process of
calculating Hamiltonians with finite parameter $t$
and eliminating their dependence on $\Delta$ by
adjusting the initial condition. More precisely,
one adjusts the counterterms so that the
coefficients of products of creation and
annihilation operators in an effective theory for
gluons of finite size $s$ become independent of
the regularization parameter $\Delta$ when the
regularization in dynamics of gluons of size zero
is being removed. The remaining unknown finite
parts must be adjusted to respect symmetries of
the theory and to match its predictions with
experiments. In the case of pure glue theory, the
only unknown parameter would be $\Lambda_{QCD}$,
which could be adjusted so that, for example, the
theory yields the desired value of mass for some
glueball, if the mass gap is found to exist.

The question arises is how the small-$x$
regularization effects can be removed. The
required counterterms are relevant to our
understanding of the theory ground state and
mechanism of confinement~\cite{Wilsonetal}. We
shall show in the next sections that the small-$x$
divergences cancel out in the third-order RGPEP
coupling constant in the effective Hamiltonians.
However, the third-order effective three-gluon
interaction terms exhibit a finite small-$x$
regularization dependence which is not yet fully
undertstood. 

One hopes that the finite small-$x$ regularization
dependence, which we illustrate using different
small-$x$ regulator factors listed in
Eqs.~(\ref{r1}) to (\ref{r3}), may cancel out in
the mass eigenvalues for glueballs and their
scattering amplitudes. Verification of such
cancellation is foreseen to be difficult because
it involves solving bound-state eigenvalue
problems for gluons. It may turn out that even for
the calculation of lightest glueballs one needs to
introduce the counterterms that also secure
confinement of color~\cite{Wilsonetal}. Before
this issue is resolved, in the practice of
approximate calculations of observables, one can
seek finite parts of the counterterms using as a
guiding rule the minimization of dependence on the
renormalization scale~\cite{PMC}, which in the
RGPEP means minimization of dependence on the
gluon size parameter $s$.

We show in the next sections that there exists a
small-$x$ regularization that yields the
third-order effective coupling constant which
depends on the size of effective gluons in the
same way as the running coupling constant
calculated using Feynman diagrams for off-shell
Green's functions depends on the virtuality of
external gluon lines. Moreover, we demonstrate
below that two different RGPEP generators lead to
the same third-order results for the effective
Hamiltonian coupling constant including finite 
effects of the small-$x$ regularization. These 
results suggest that the calculated finite 
sensitivity to the small-$x$ regularization is 
not accidental and, being established here, 
should be further studied as a potentially 
universal feature of a whole class of FF 
Hamiltonians for effective gluons.

\section{ Calculation of the three-gluon term }
\label{ThreeGluonTerm}

We solve the RGPEP Eq.~(\ref{H'text}) pertubatively, 
expanding $H_t$ in powers of the coupling 
constant $g$ up to third order,
\beq
\label{Hpert}
H_t \es 
H_{11,0,t} + H_{11,g^2,t} + H_{21,g,t} + 
H_{12,g,t} + H_{31,g^2,t} + H_{13,g^2,t} + 
H_{22,g^2,t} + H_{21,g^3,t} + H_{12,g^3,t}  \ .
\eeq
The first subscript lists the numbers of creation 
and annihilation operators contained in a term,
correspondingly. The second subscript indicates 
the order in powers of $g$, and the third subscript 
indicates dependence on the parameter $t$. For  
building intuition, we introduce symbols: $\mu^2$ 
for mass terms, which have the first subscript 11;
$Y$ for three-gluon interaction terms, which have 
the first subscripts 12 or 21; and $X$ for four-gluon 
interaction terms, which have the first subscript 22. 
Consequently, the powers of $g$ are explicitly accounted 
for using the following notation:
\beq
H_{11,0} & \to & E \ , \\
H_{11,g^2} & \to & g^2 \hat \mu^2 \ , \\
H_{21,g} + H_{12,g} & \to & g Y_{21} + g Y_{12} \ , \\
H_{22,g^2} & \to & g^2 X_{22} \ , \\
H_{31,g^2} + H_{13,g^2} & \to & g^2 \Xi_{31} + g^2 \Xi_{13} \ , \\ 
H_{21,g^3} + H_{12,g^3}  & \to & g^3 Y_{h21} +
g^3 Y_{h12} \ .
\eeq
To be faithful to the difference in notation between
$\cH_t$ and $H_t$, associated with changing bare to 
effective gluon operators, see Appendix~\ref{AppRGPEP}, 
we could also introduce symbols like $\cE$ instead of $E$,
$\cY_t$ instead of $Y_t$, etc. However, it is simpler to 
remember the difference, and write 
\beq
H_t \es 
E + g^2 \hat \mu^2_t + g Y_{21t} + g Y_{12t} + g^2 X_{22t}
+ g^2 \Xi_{31t} + g^2 \Xi_{13t} + 
g^3 Y_{h21t} + g^3 Y_{h12t} \ .
\label{Ht}
\eeq
The initial condition Hamiltonian at $t=0$, 
is written using symbols with the subscript 
0 in place of $t$,
\beq
H_0 \es 
E + g^2 \hat \mu_0^2 + g Y_{210} + g Y_{120} + g^2 X_{220}
+ g^2 \Xi_{310} + g^2 \Xi_{130} + 
g^3  Y_{h210} + g^3  Y_{h120} \ .
\eeq
The counterterms that need to be found are included in $H_0$. 

Besides using the parameter $t$ and its initial value $t=0$, 
we also use the parameter $\lambda = s^{-1} = t^{-1/4}$, 
whose initial value is $\infty$. The parameter $\lambda$ has 
the interpretation of momentum-space width of the form 
factors that appear in solutions for $H_t$.

The RGPEP Eq.~(\ref{H'text}) for the
Hamiltonian~(\ref{Ht}) reads
\beq
& & 
g^2  \partial_t \hat \mu_t^2 + g Y_{21t}' + g Y_{12t}' + g^2 X_{22t}'
+ g^2 \Xi_{31t}' + g^2 \Xi_{13t}' + 
g^3 Y_{h21t}' + g^3 Y_{h12t}' \nn
\es
\left[ \left[ E ,
g Y_{21Pt} + g Y_{12Pt} + g^2 X_{22Pt}
+ g^2 \Xi_{31Pt} + g^2 \Xi_{13Pt} + 
g^3 Y_{h21Pt} + g^3 Y_{h12Pt} \right], 
\right. 
\nt
\left.
E + g^2 \hat \mu^2_t + g Y_{21t} + g Y_{12t} + g^2 X_t
+ g^2 \Xi_{31t} + g^2 \Xi_{13t} + 
g^3 Y_{h21t} + g^3 Y_{h12t} \right] \ . 
\label{RGPEPeqGluons}
\eeq 
We solve Eq.~(\ref{RGPEPeqGluons}) order-by-order
in series of powers of $g$, which eventually is 
translated into a series expansion in powers of 
$g_t$. Mass squared terms are of the second order and 
the gluon vertex is made of terms of the first and 
third order. After removing powers of $g$ from the 
equations,
\beq
Y_{21t}' + Y_{12t}' 
\es
\left[ \left[ E , Y_{21Pt} + Y_{12Pt} \right], E \right] \ ,\label{RGeq1st} \\
\partial_t \hat \mu^2_t + X_{22t}' + \Xi_{31t}' + \Xi_{13t}'  
\es
\left[ \left[ E , X_{22Pt} + \Xi_{31Pt} + \Xi_{13Pt} \right], E \right] 
\np
\left[ \left[ E , Y_{21Pt} + Y_{12Pt}\right], Y_{21t} + Y_{12t} \right] \ , \label{RGeq2nd} \\
Y_{h21t}' + Y_{h12t}'
\es
\left[ \left[ E , Y_{h21Pt} + Y_{h12Pt} \right], E  \right] 
\np 
\left[ \left[ E , X_{22Pt} + \Xi_{31Pt} + \Xi_{13Pt} \right], 
Y_{21t} + Y_{12t} \right] 
\np
\left[ \left[ E , Y_{21Pt} + Y_{12Pt} \right], 
\hat \mu^2_t + X_{22t} + \Xi_{31t} + \Xi_{13t} \right] \ .
\label{RGeq3rd}
\eeq
The running of the Hamiltonian coupling constant $g_t$
is encoded in the operator Eq.~(\ref{RGeq3rd}). 
Solving Eq.~(\ref{RGeq3rd}) requires knowledge of
solutions to the operator Eqs.~(\ref{RGeq1st}) and 
(\ref{RGeq2nd}). The gluon mass squared counterterm 
is obtained from the second-order equations, and 
the three-gluon vertex counterterm from the 
third-order equations. 

The solution for 
\beq
\label{Ytext}
Y_t \es g (Y_{12t} + Y_{21t}) + g^3 (Y_{h21t} 
+ Y_{h12t}) \ ,
\eeq 
is written in terms of the bare creation and 
annihilation operators for canonical gluons 
and powers of the bare coupling constant. The 
last step in the RGPEP is the replacement of 
the bare gluon operators by the effective ones 
at scale $t$ and expressing $g$ in terms of $g_t$. 

Details of solving Eqs.~(\ref{RGeq1st}) to (\ref{RGeq3rd}) 
are described in Appendix \ref{RGPEPsolution}. Here we 
list the results. The first-order terms are the same as 
in Ref.~\cite{Glazek:2000dc}, see Appendix \ref{1storderAppendix}.

\subsection{ Mass squared term and its counterterm} 
\label{massct}

The second-order mass squared term for effective 
gluons that solves Eq.~(\ref{RGeq2nd}), has the form
\beq
\label{hatmubare}
\hat\mu^2_t = \sum_{\sigma c}\int [k] \ {\mu^2_t
\over k^+} \ a^\dagger_{k \sigma c} a_{k \sigma c} \ ,
\eeq
where the only element that depends on the scale 
$t$ is the parameter $\mu^2_t$. The result for it 
reads
\beq
\mu^2_t \es \mu^2_\delta + {g^2 \over (4\pi)^2 } \int_0^1 dx \,
r_{\delta \mu}(x) \sum_{12}
|Y_{12k}|^2/\kappa^2 \int_0^\infty dz \, \exp{(-2tz^2)} \ , 
\label{gluonmasssol}
\eeq
where 
\beq
\sum_{12}
|Y_{12k}|^2 / \kappa^2
\es 
N_c [1 + 1/x^2 + 1/(1-x)^2] 
\rs 
P(x)/[2x(1-x)] \ , 
\eeq
and $P(x)$ is the Altarelli-Parisi gluon splitting
function $P_{GG}(x)$~\cite{AP}. $N_c = 3$ denotes
the number of colors. The effective mass squared
term is sensitive to the small-$x$ regularization.
The counterterm that canceled dependence on the
ultraviolet cutoff $\Delta\to\infty$ contains 
the mass-squared factor of the form
\beq
\mu^2_0 \es 
\mu^2_\delta + {g^2 \over (4\pi)^2 } \int_0^1 dx \,
r_{\delta \mu}(x) P(x) \int_0^\infty dz \, \exp{[-4z x(1-x)/\Delta^2]} \ .
\eeq
Comparison with Ref.~\cite{Glazek:2000dc} shows 
that the gluon mass-squared term obtained using 
the RGPEP generator of Eq.~(\ref{cG1}) does not 
differ from the one obtained using the generator 
of Eq.~(\ref{cG2}).

The ultra-violet finite part of the mass
counterterm, $\mu^2_\delta$, depends on the
small-$x$ regularization parameter $\delta$ in the
initial Hamiltonian. Therefore, the simplest way
of choosing the ultra-violet finite part of the
mass-squared counterterm is to set the mass
squared for effective gluons at some value of $t$
to a desired function of $\delta$. Such function
of $\delta$ can be fixed by demanding that the
effective Hamiltonian eigenvalue for lightest
states with color quantum numbers of a single
gluon contains a specified mass-squared term that
depends on the parameter $\delta$ in a specific
way. 

The right dependence for defining
a complete theory of gluons is currently unknown 
but it is also currently not excluded that one 
can attempt to describe confinement of gluons 
by demanding that the gluon mass eigenvalue 
diverges in the limit $\delta \to 0$. Verification 
of this option requires studies far beyond the 
scope of this article. Namely, one needs to study 
terms of higher order and consider the eigenvalue 
problem in higher order than third, before one 
will know if the perturbative expansion of the 
RGPEP can lead to establishment of a general
structure of the Hamiltonian that may solve 
Eq.~(\ref{H'text}) beyond perturbation theory. 
Here, we shall find that, once the ultraviolet 
divergences are removed, the small-$x$ divergences 
do not appear in the third-order asymptotically 
free coupling constant in renormalized Hamiltonians 
for effective gluons.

The second-order effective gluon mass term, denoted 
by $\hat m^2_t$, is obtained from $\hat \mu^2_t$ in 
Eq.~(\ref{hatmubare}) by applying the transformation 
$\cU_t$ and thus replacing the creation and annihilation 
operators for bare gluons by the ones for effective 
gluons of size $s$. Namely,
\beq
\label{hatmueff}
\hat m^2_t \es \cU_t \ \hat\mu^2_t \ \cU_t^\dagger \ .
\eeq
This transformation amounts to the replacement in 
Eq.~(\ref{hatmubare}) of $a^\dagger_{k \sigma c} 
a_{k \sigma c}$ by $a^\dagger_{t \, k \sigma c} 
a_{t \, k \sigma c}$. The former operators correspond
to thin and the latter to thick lines in Fig.~\ref{Figgluonmass}.

\subsection{ Third-order three-gluon term and its counterterm } 
\label{gvct}

We focus our attention on the term $H_{21,g^3,t}$
in Eq.~(\ref{Hpert}), knowing that $H_{12,g^3,t}$
is its Hermitian conjugate. The term has the 
structure  
\beq
 H_{21,g^3,t} =  {\cal U}_t \ \gamma_{t\, 21} \ {\cal U}^\dagger_t \ ,
\eeq
where
\beq
\gamma_{t\, 21}= f_t \sum_n \gamma_{t\,21(n)} \ .
\eeq
The factor $f_t$ in front of an operator means that 
the vertex functions in the operator are multiplied 
by the form factor defined in Eq.~(\ref{f}).

The subscript $n$ in the sum ranges over ten 
values, denoted by alphabet letters from $a$ 
to $j$. Each of the summed terms results from 
some specific operator product in Eq.~(\ref{RGeq3rd}). 
In each of these terms there appears a vertex 
function, denoted by $\gamma_{(n)}$, in the otherwise 
universal pattern of the formula
\beq
\gamma_{t\, 21(n)} \es  \sum_{123}\int[123] \ \tilde \delta(k_1 + k_2 -
k_3) \ { g^3 \over 16 \pi^3} \,\, {1 \over 2} \,\, \gamma_{(n)}
\ a^\dagger_1 a^\dagger_2 a_3 \ .  
\eeq
The vertex functions $\gamma_{(n)}$, with
subscript $n$ ranging from $a$ to $j$, are given
in Appendix~\ref{3rdorder}, with their operator
origin in Eq.~(\ref{RGeq3rd}) being illustrated by
diagrams in Fig.~\ref{Fig3gluonvertex} there. The
thick external lines in Fig.~\ref{Fig3gluonvertex}
correspond to the creation and annihilation
operators that appear in the three-gluon
interaction term for effective gluons of size $s$.
The thin internal lines correspond to the
commutators that result from moving annihilation
operators to the right of all creation operators
for gluons of size zero, in terms of which the
RGPEP Eq.~(\ref{H'text}) is solved. Lines with a
transverse dash indicate instantaneous
interactions in $H_0$ that result from the
constraint of Eq.~(\ref{A-1}).

The vertex functions diverge when the ultra-violet
cutoff parameter $\Delta$ is being sent to
infinity. The divergences result from integration
over the transverse relative momentum of virtual
quanta whose creation and annihilation operators
were contracted in the products that appear on the
right-hand side of Eq.~(\ref{RGeq3rd}), cf.
Eq.~(\ref{Yh21t}). The required counterterm,
calculated in Appendix~\ref{threegCT}, has the
form
\beq
\tilde Y_{h210} \es  \sum_{123}\int[123] \ \tilde \delta(k_1 + k_2 -
k_3)\,\, { g^3 \over 16 \pi^3} \,\, \gamma_0 \,\, a^\dagger_1
a^\dagger_2 a_3 \ ,
\eeq
with the vertex function $\gamma_0$ derived in 
Eq.~(\ref{gammainfty}),
\beq
\label{gamma0+finite}
\gamma_0 =- Y_{123} { \pi \over 3} \, \ln{ \Delta \over \mu}
\left\{ N_c\left[ 11 + h(x_1) \right] \right\} +
\gamma_{\text{finite}} \ .
\eeq
For calculation of the Hamiltonian running coupling 
constant, the finite part of the counterterm, denoted 
in Eq.~(\ref{gamma0+finite}) by $\gamma_{\text{finite}}$, 
will not need to be specified when the subtraction of 
the diverging part is introduced as described in the 
next section.
 
\subsection{ Running coupling constant } 
\label{runningcoupling}

Our Hamiltonian running coupling constant $g_t$ is
extracted from the three-gluon terms in $H_t$.
These terms create two gluons and annihilate one,
or vice versa. Both types yield the same result
for $g_t$. The three-gluon term is a sum of terms
denoted by ($a$) to ($j$) in the previous section.
The vertex function of the entire sum depends on 
the gluon colors, polarizations and momenta. In
the terms that vary with the scale parameter $t$, 
the dependence on color and polarization in the 
limit $\kappa_{12}^\perp \to 0$ takes the form of 
a combination $Y_{123}$ shown in Appendix~\ref{initialH} 
in Eq.~(\ref{Y123}). This combination is multiplied 
by a function $g(t,x_1)$, where $x_1 = 1-x_2$ refers 
to the $+$-momentum fraction carried by one gluon 
of the total momentum of two gluons that are created 
or annihilated by the three-gluon term. The coupling 
constant $g_t$ is defined as the value of $g(t,x_1)$ 
at some value of $x_1 = x_0$,
\beq
g_t \es g(t,x_0) \ .
\eeq 
Note that the Hamiltonian $\cH_t$ that appears in 
Eq.~(\ref{H'text}) is calculated using the bare 
creation and annihilation operators and the 
effective Hamiltonian $H_t$ is obtained from $\cH_t$ 
by inserting the effective creation and annihilation 
operators in place of the bare ones. The vertex
function is not changed. We calculate $g_t$ using
$\cH_t$.

The counterterm contribution to the vertex function
can be written as
\beq
\label{gamma0t0}
\gamma_0 \es - \gamma_{t_0} 
+ \tilde \gamma_{\text{finite}} \ .
\eeq
The diverging part of the counterterm is thus 
specified using the negative of $\gamma_t$ at 
an arbitrary finite value of $t_0$. Therefore, 
$\tilde \gamma_{\text{finite}}$ may differ from 
$\gamma_{\text{finite}}$ in Eq.~(\ref{gamma0+finite}) 
by terms that do not depend on $t$. Such terms
will not contribute to the dependence of $g_t$ 
on $t$ and will not be further discussed in this 
article. Note that since the diverging part of 
the counterterm may be a function of $x_1$, as
displayed in Eq.~(\ref{gamma0+finite}), one also 
has to consider the counterterm finite part that 
may be a function of $x_1$~\cite{Wilsonetal}.

After inclusion of the counterterm defined in
Eq.~(\ref{gamma0t0}), our result for the three-gluon 
interaction term in $\cH_t$ has the form (the 
symbol $\sigma$ stands for spin variables)
\beq
Y_t \es g Y_{1t} + g^3 Y_{3t} + ... \ , \\
Y_t
\es
\sum_{123}\int[123] \  \tilde \delta(1+2-3) 
\ f_{12t} \ \tilde Y_t(x_1,\kappa_{12}^\perp, \sigma)  \ a^\dagger_1 a^\dagger_2 a_3   \ , \\
Y_{1t}
\es
\sum_{123}\int[123] \  \tilde \delta(1+2-3) 
\ f_{12t} \ \tilde Y_{1t}(x_1,\kappa_{12}^\perp, \sigma)  \ a^\dagger_1 a^\dagger_2 a_3   \ , \\
Y_{3t}
\es
\sum_{123}\int[123] \  \tilde \delta(1+2-3) 
\ f_{12t} \ \tilde Y_{3t}(x_1,\kappa_{12}^\perp, \sigma)  \ a^\dagger_1 a^\dagger_2 a_3   \ .
\eeq
The symbols $\tilde Y$ denote vertex
functions without the form factor $f_t$.
Assuming a counterterm that involves a 
subtraction at some $t=t_0$ as described
above, one obtains the third-order vertex 
factor $\tilde Y$ of the structure 
\beq
\tilde Y_{3t}(x_1,\kappa_{12}^\perp, \sigma) 
\es 
       \tilde T_{3t}(x_1,\kappa_{12}^\perp, \sigma)  
     - \tilde T_{3t_0}(x_1,\kappa_{12}^\perp, \sigma) 
     + \tilde T_{3 \,
\text{finite}}(x_1,\kappa_{12}^\perp, \sigma)  \ ,
\eeq 
where the symbol $\tilde T$ denotes the sum of third-order terms
from ($a$) to ($i$) calculated in Appendix \ref{3rdorder},
\beq
\tilde T \es \sum_{n= \, a}^{i} \  \gamma_{(n)} \ .
\eeq
The diverging part of the counterterm denoted as 
$\gamma_{(j)}$ in Eq.~(\ref{gammajapp}) in App. \ref{3rdorder}, 
is included here through the subtraction at $t=t_0$, and
the associated change in the finite part is not needed
in the discussion that follows. 

Combined, all the three-gluon terms in expansion 
up to third-order in powers of $g$, have the form
\beq
\tilde Y_t(x_1,\kappa_{12}^\perp, \sigma) 
\es g \tilde Y_{1t}(x_1,\kappa_{12}^\perp, \sigma)  
+ g^3 \left[ \tilde T_{3 t }(x_1,\kappa_{12}^\perp, \sigma)  
           - \tilde T_{3 t_0}(x_1,\kappa_{12}^\perp, \sigma) 
           + \tilde T_{3 \text{finite}}(x_1,\kappa_{12}^\perp, \sigma)  \right] \ .
\eeq
By our definition, the Hamiltonian coupling constant 
$g_t$ is found as a coefficient in front of the canonical 
color, spin and momentum dependent factor $Y_{123}(x_1,\kappa_{12}^\perp, 
\sigma)$ of Eq.~(\ref{Y123}), in the limit $\kappa_{12}^\perp
\to 0$, for some value of $x_1$, denoted by $x_0$. 
At some arbitrary value of $t=t_0$, $g_{t_0}$
must be set to a specific finite value $g_0$ 
that produces agreement with experiment when 
one describes data using the Hamiltonian with 
$t = t_0$. 

We obtain
\beq
\label{consider}
\lim_{\kappa_{12}^\perp \to 0}
\tilde Y_t(x_1,\kappa_{12}^\perp, \sigma) 
\es
\lim_{\kappa_{12}^\perp \to 0}
\left[
c_t(x_1,\kappa_{12}^\perp) 
Y_{123}(x_1,\kappa_{12}^\perp, \sigma) 
+ 
g^3 \tilde T_{3 \,\text{finite}}(x_1,\kappa_{12}^\perp, \sigma) 
\right]  
\\
\es 
\lim_{\kappa_{12}^\perp \to 0}
g Y_{123}(x_1,\kappa_{12}^\perp, \sigma)
\np
\lim_{\kappa_{12}^\perp \to 0} g^3 
\left[
  c_{3 t}(x_1,\kappa_{12}^\perp) 
- c_{3 t_0 }(x_1,\kappa_{12}^\perp) 
\right] 
Y_{123}(x_1,\kappa_{12}^\perp, \sigma) 
\np
\lim_{\kappa_{12}^\perp \to 0}
g^3 \, \tilde T_{3 \,\text{finite}}(x_1,\kappa_{12}^\perp, \sigma) 
\ .
\eeq
Removing $Y_{123}(x_1,\kappa_{12}^\perp, \sigma)$
from all terms besides the term $\tilde T_3$ that 
does not have to have the spin and momentum 
structure of $Y_{123}(x_1,\kappa_{12}^\perp, \sigma)$, 
in the limit,
\beq
\lim_{\kappa_{12}^\perp \to 0}
c_t(x_1,\kappa_{12}^\perp) 
\es
g + g^3 \lim_{\kappa_{12}^\perp \to 0}
\left[
  c_{3t  }(x_1,\kappa_{12}^\perp) 
- 
  c_{3t_0}(x_1,\kappa_{12}^\perp) 
\right] \ , 
  \label{limitc3}
\eeq
or 
\beq 
\label{limitc4}
c_t(x_1,0^\perp) 
\es 
g + g^3 
\left[
c_{3t  }(x_1,0^\perp) 
- 
c_{3t_0}(x_1,0^\perp) 
\right] \ .
\eeq
Evaluation of the limit in Eq.~(\ref{limitc3}) 
that yields Eq.~(\ref{limitc4}) is described 
in Appendix~\ref{Limitk12}.

Setting $x_1 = x_0$ and dropping the argument 
$\kappa_{12}^\perp$ set to zero, our definition 
of the coupling constant $g_t$ leads to
\beq
\label{gl1}
g_t
& \equiv &  c_t(x_0) \\
\es 
\label{gl2}
g + g^3 
\left[
  c_{3t  }(x_0) 
- 
  c_{3t_0}(x_0) 
\right] \ .
\eeq
We calculate the coupling constant $g$ that appears 
in the initial Hamiltonian after inclusion of the 
counterterm, by demanding that at $t=t_0$ the coupling 
constant should have the value $g_0$,
\beq
g_{t_0} \es g_0 \ .
\eeq
The value of $g_0$ is determined by comparison with data 
using the Hamiltonian corresponding to $t = t_0$. Hence,
with accuracy to terms of order $g_0^3$ or smaller,
Eq.~(\ref{gl2}) implies
\beq
\label{gl3}
g_t 
\es 
g_0 + g_0^3 
\left[
  c_{3t  }(x_0) 
- 
  c_{3t_0}(x_0) 
\right] \ .
\label{gtx0}
\eeq
The right-hand side of this result is calculated
in Appendix~\ref{Limitk12}. It can also be expressed 
as a function of momentum scale $\lambda=1/s$, which 
facilitates comparison with Refs.~\cite{Gross:1973id,
Politzer:1973fx} and~\cite{Glazek:2000dc}. For this
purpose, we denote $g_t$ by $g_\lambda$ when we set 
$t = \lambda^{-4}$.

From all terms that contribute to the right-hand 
side of Eq.~(\ref{gl3}), listed as ($a$) to ($i$) 
in Appendix~\ref{Limitk12}, only the contributions 
of $\gamma_{(a)}$, $\gamma_{(d)}$ and $\gamma_{(g)}$ 
are different from zero. These terms yield
\beq
\label{gts}
g_\lambda \es
g_0 - { g^3_0 \over 48 \pi^2 }   N_c \,  \left[ 11 + h(x_0) \right]\,\ln {\lambda \over \lambda_0} \ ,
\eeq 
where 
\beq
h (x_0) \es \chi(x_0) + \chi(1-x_0) \ , \\
\chi (x_0) 
\es
6 \int_{x_0}^1 dx \  r_{\delta Y} \left[ 2/(1-x) + 1/(x-x_0) + 1/x \right] \ - \
9 \ \tilde r_{\delta} (x_0) \int_0^1dx\,r_{\delta \mu}(x)  \, \left[ {1 \over x} + { 1 \over 1 - x}\right] \ .
\eeq
This result depends in a finite way 
on our regularization of small-$x$ 
divergences.

\begin{figure}[h]
\hspace{0.2cm}
\includegraphics[width=0.44\textwidth]{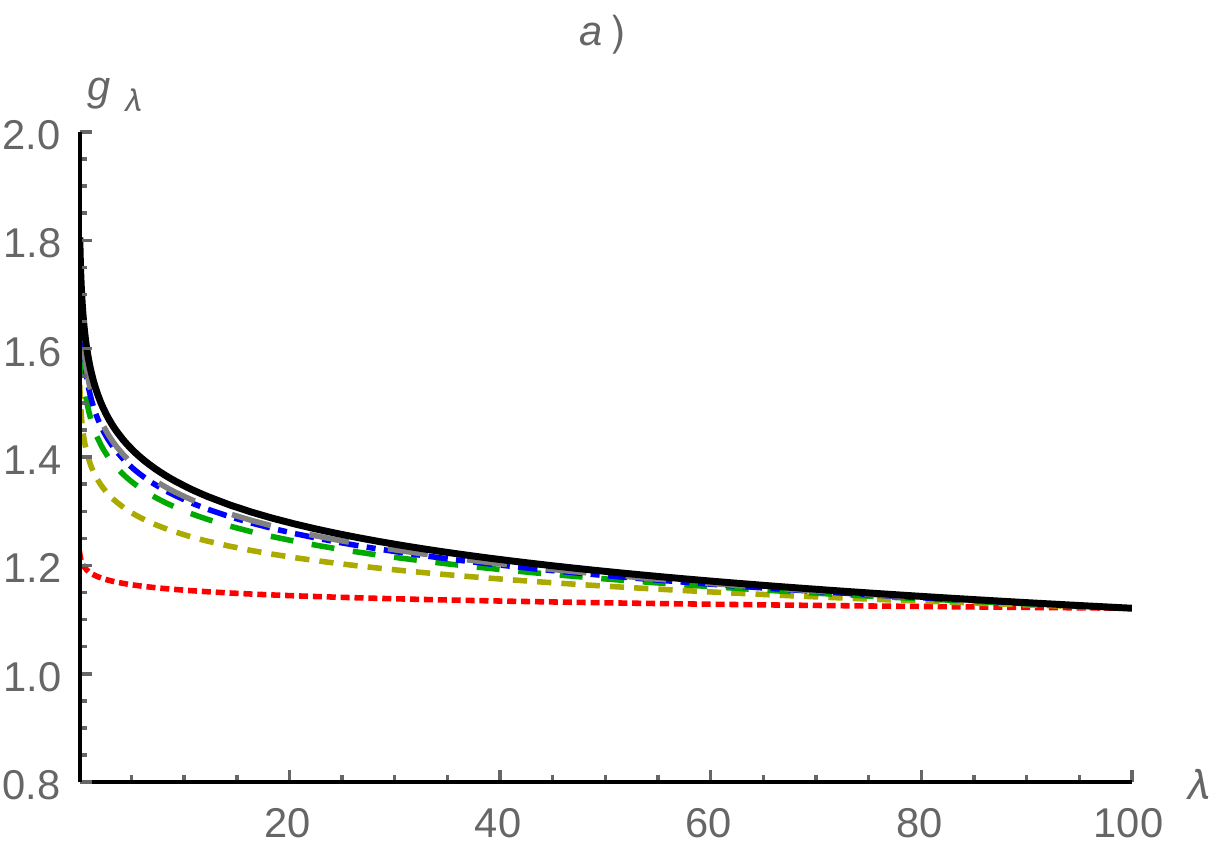}
\hspace{0.2cm}
\includegraphics[width=0.49\textwidth]{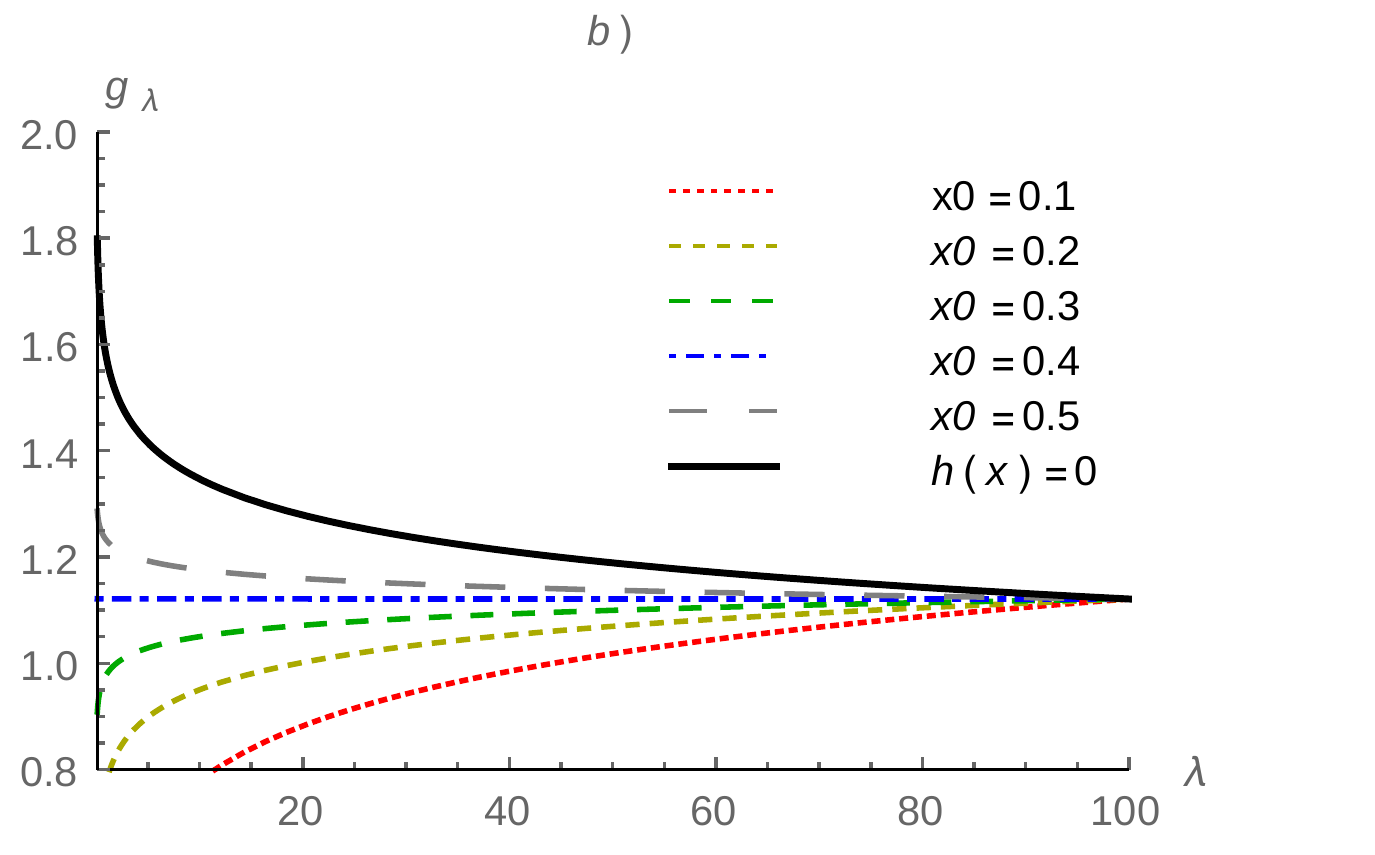}
\caption{\label{Vx}
The FF Hamiltonian third-order RGPEP running 
coupling constant for effective gluons, 
$g_\lambda$ of Eq.~(\ref{gts}), is drawn using 
different dashed lines for different values of 
$x_0$, as a function of $\lambda$ in GeV, starting 
from an arbitrarily chosen value $g_0=1.1$ at 
$\lambda_0 = 100$ GeV. Plots a) and b) correspond 
to small-$x$ regularizations in Eqs.~(\ref{r1})
and~(\ref{r2}). The thick continuous lines in
both plots show one and the same result for the 
regularization in Eq.~(\ref{r3}), which exhibits 
no dependence on $x_0$.}
\end{figure}

For the choices of small-$x$ regularization 
that are listed in Sec.~\ref{reg} in Eqs.~(\ref{r1}), 
(\ref{r2}) and (\ref{r3}), to which we refer as 
versions a), b) and c), the limit $\delta \to 0$ 
yields the Hamiltonian running coupling with 
the function $h(x)$ given by, correspondingly,
\beq
\text a ) & & \quad  h(x_0) \, = \, 12\left[ 3 + {1-x_0 - x_0^2 \over (1-x_0)(1-2x_0)} \ln{x_0} +
                               {(1-x_0)^2 -x_0 \over x_0(1-2x_0) } \ln{(1-x_0)} \right] \ , \\
\text b ) & & \quad  h(x_0) \, = \, 12 \ln{ \, \text{min}(x_0,1-x_0)} \ , \\
\text c ) & & \quad  h(x_0) \, = \, 0 \  . 
\eeq
The running of the Hamiltonian coupling constant 
described by Eq.~(\ref{gts}) for small-$x$ 
regularizations a) and b), is illustrated by the 
dashed curves in plots a) and b) of Fig.~\ref{Vx}, 
respectively. Different dashed curves correspond 
to different values of $x_0$. Only examples with 
$x_0$ between 0.1 and 0.5 are plotted, because the 
function $h(x_0)$ in Eq.~(\ref{gts}) is symmetric 
with respect to the change $x_0 \rightarrow 1-x_0$. For 
the regularization c), we have $h(x_0) = 0$ irrespective 
of the value of $x_0$, and
\beq
\label{gl}
g_\lambda \es
g_0 - { g_0^3 \over 48 \pi^2 }   N_c \,   11 \,\ln
{ \lambda \over \lambda_0} \ ,
\eeq 
which is illustrated by one and the same continuous 
line in both plots a) and b) of Fig.~\ref{Vx}. 
Differentiation of Eq.~(\ref{gl}) with respect 
to $\lambda$ produces
\beq
\lambda {d \over d\lambda} \, g_\lambda
\es \beta_0 g_\lambda^3 \ ,
\eeq
where 
\begin{equation}
 \beta_0 \rs - { 11 N_c \over 48\pi^2 } \ .
\end{equation}
This result matches the asymptotic freedom result 
in Refs.~\cite{Gross:1973id,Politzer:1973fx},
when one identifies $\lambda$ with the momentum 
scale of external gluon lines in Feynman diagrams.
Discussion of this result is provided below in 
Sec.~\ref{Universality}.

\section{ Universality of the RGPEP solution } 
\label{Universality}

As a result of the third-order RGPEP, the three-gluon 
interaction term in effective FF Hamiltonians for 
gluons is
\beq
\label{H3t}
H_{t A^3} \es \sum_{123}\int[123] \, \tilde \delta(p^\dagger - p)
\ f_{12t} \ \tilde Y_t(x_1,\kappa_{12}^\perp, \sigma) 
\ a^\dagger_{t1} a^\dagger_{t2} a_{t3} +
H. c.  \ ,
\eeq
where with accuracy to terms order $g_0^3$ one has
\beq
\tilde Y_t(x_1,\kappa_{12}^\perp, \sigma) 
\es g_0 \, Y_{123}(x_1,\kappa_{12}^\perp, \sigma)  
+   g_0^3 \left[ \tilde T_{3 t }(x_1,\kappa_{12}^\perp, \sigma)  
           - \tilde T_{3 t_0}(x_1,\kappa_{12}^\perp, \sigma) 
           + \tilde T_{3 \text{finite}}(x_1,\kappa_{12}^\perp, \sigma)  \right]
\ .
\eeq
For infinitesimal $\kappa_{12}^\perp$,
\beq
\tilde Y_t(x_1,\kappa_{12}^\perp, \sigma) 
\es 
V_t(x_1) \ Y_{123}(x_1,\kappa_{12}^\perp, \sigma)   
+ 
g_0^3 \tilde T_{3 \text{finite}}(x_1,\kappa_{12}^\perp, \sigma) 
+
o(\kappa_{12}^{\perp}) \ ,
\eeq
where
\beq
V_t(x_1) 
\es
g_t 
+ 
g_t^3 \left[ c_{3 t }(x_1)  - c_{3 t }(x_0) 
           - c_{3 t_0}(x_1) + c_{3 t_0}(x_0) \right] \ ,
\eeq
$g_t$ is given in Eq.~(\ref{gl3}) and the coefficients
$c_3$ are described in Sec.~\ref{runningcoupling}.
Universality of this result is claimed 
on the basis of comparison with results
obtained in Refs.~\cite{Gross:1973id,
Politzer:1973fx} and~\cite{Glazek:2000dc}.

Comparison with Refs.~\cite{Gross:1973id,
Politzer:1973fx} shows that the FF Hamiltonian
running coupling constant $g_\lambda$ exhibits, 
in the RGPEP of third order, the same leading
dependence on the momentum width of vertex form
factor $\lambda$, as the running coupling constant
in Refs.~\cite{Gross:1973id,Politzer:1973fx}
exhibits as a function of the length $\lambda$ of
Euclidean momenta of external gluon lines in the
three-point effective action. In order to compare
these two results, one has to assume that the
Euclidean Green's functions correspond, by some
continuation procedure from imaginary to real time
variable, to a Minkowski space-time quantum theory
in which a renormalized Hamiltonian has a
three-gluon interaction term of a specific
dependence on the momentum scale parameter
$\lambda$. Our calculation suggests, but does not
prove, that the Euclidean scale $\lambda$
corresponds to the RGPEP width $\lambda$. Namely,
the FF Hamiltonian matrix element that appears in
the virtual transition amplitude between one- and
two-gluon states in the Fock space of effective
gluons, is suggested to correspond to the
continuation of the three-point Eculidean Green's
function, or effective action, to the Minkowski
variables. The suggestion is not verifiable by any
simple continuation because we do not fully know
the analytic structure of either function.
However, the observed universality of asymptotic
freedom in both the perturbative Euclidean Green's
function calculus and Minkowskian Hamiltonian
quantum mechanical operator calculus points out a
direction in which one can seek a constructive
demonstration that these two ways of defining a
theory are equivalent.

Comparison with Ref.~\cite{Glazek:2000dc},
where the Hamiltonian three-gluon vertex 
is calculated as a function of the momentum 
scale $\lambda$ using the RGPEP with a different 
generator than the one used here, is facilitated 
by observing that the size of effective gluons, 
$s$, is equal to the inverse of $\lambda$. 
Using this relation, one sees that the present 
result is the same as in~\cite{Glazek:2000dc}, 
despite that the generators are different. 
Specifically, using $\cG_t$ in Eq.~(\ref{cG2}) 
instead of the one in Eq.~(\ref{cG1}) does not 
change the third-order results for $g_\lambda$. 
Fig.~\ref{Vx} illustrates this finding by showing 
the present results for $g_\lambda$. The
current calculation explicitly extends the 
universality of leading perturbative terms in 
the beta-function to the RGPEP calculus for 
Hamiltonian operators in the effective particle 
Fock space. 

Thus, the universality we claim is two-fold.
One universal aspect is that the third-order
RGPEP Hamiltonian running coupling constants 
exhibit the same asymptotic freedom behavior 
that is known to be universal in the calculus 
based on the renormalized Feynman diagrams. 
This is of interest from the point of view 
that the Hamiltonian quantum mechanics in the 
Minkowski space-time and the Feynman diagrams 
for virtual transition amplitudes can be precisely 
related to each other in a relativistic theory
including renormalization, which generally 
remains to be desired~\cite{Wilson:1965zzb,
Wilson:1970tp}. Such relation is needed for 
incorporating non-perturbative features of 
hadrons in calculations that so far remain 
limited to the usage of qualitative and quantitative 
input from the parton model.

The other universal aspect is that the third-order
Hamiltonian running coupling constant depends on
the size of effective gluons, or momentum width of
effective Hamiltonian three-gluon vertex, in a way
that does not depend on the choice of the RGPEP
generator. This is of great interest in view of
the fact that the presently used generator does
not depend on the derivative of the Hamiltonian
with respect to $t$, while the previously used
generator does. Hence, the obtained stability of
asymptotically free behavior of effective gluon
interactions, with respect to change of the RGPEP
generator, suggests a viable way around the
difficult problem of solving for the derivative of
the Hamiltonian in terms of the Hamiltonian
itself. The generator used here is thus shown to
offer a way of seeking non-perturbative solutions
to the RGPEP equation in a greatly simplified
setup in comparison with the original one.

\section{ Conclusion }
\label{Conclusion}

Knowledge of the third-order RGPEP result
for the Hamiltonian of effective gluons is
not sufficient for setting up any physical 
eigenvalue problem, such as the eigenvalue 
problem for a glueball. At least fourth-order
terms are needed, which describe interactions
among two effective gluons including the 
effect of running of the coupling constant.
Such calculations are considerably more 
involved than the third-order calculations
described here. 

However, the presently used RGPEP generator, 
demonstrated here to imply the Hamiltonian 
running coupling constant of the form that 
is familiar from other formalisms and 
renormalization schemes, turns out to lead 
to a considerably simpler third-order calculation 
than the previously used generator did. The consequence 
of this result is that the required fourth-order 
calculations with the presently used generator 
are expected to be considerably simpler than 
they could have been with the previously used 
generator. 

In particular, the same finite effects of
small-$x$ regularization are found using
the present generator and the previous one. 
Since the FF Hamiltonian mechanisms of 
confinement and chiral symmetry breaking 
are expected to be related to the gluon 
dynamics at small-$x$, the simpler generator 
than the one used before is welcome as a tool 
for studying the small-$x$ dynamics in full 
QCD.

\vskip.2in
\centerline{\bf  Acknowledgments }
\vskip.1in

MGR acknowledges financial support from 
the Austrian Science Fund (FWF), project 
nr. P25121-N27, the Institute of High Energy Physics 
of the \"OAW and the Paul Urban Fund, and
she thanks members of the Institute of 
Theoretical Physics at the University 
of Warsaw Physics Department, where a 
part of this work was done, for hospitality.

\begin{appendix}

\section{ Details of the RGPEP }
\label{AppRGPEP}

The effective Hamiltonian is related to the
regulated canonical one with counterterms by 
the condition of no dependence on the arbitrary 
RGPEP scale parameter $t$,
\beq
\cH_t(a_t) \es \cH_0(a_0)  \ .
\eeq
This condition implies via Eq.~(\ref{at}) that
\beq
\label{cHt}
\cH_t(a_0) = \cU_t^\dagger \cH_0(a_0)\cU_t  \ .
\eeq
Differentiation of Eq.~(\ref{cHt}) with respect 
to $t$ yields
\beq 
\label{ht1}
\cH'_t(a_0) \es
[ \cG_t(a_0) , \cH_t(a_0) ] \ ,
\eeq 
where $\cG_t = - \cU_t^\dagger \cU'_t$ is called 
the RGPEP generator and the related solution for 
$\cU_t$ is given in Eq.~(\ref{Usolution}).

We consider two different generators, one from
Ref.~\cite{Glazek:2000dc},
\beq
\label{cG1}
\cG_t \es \left\{ (1-f_t^{-1})\cH_t \right\}_{\cH_f}\ ,
\eeq
and another one from Ref.~\cite{Glazek:2012qj},
\beq
\label{cG2}
\cG_t \es [ \cH_f, \cH_{Pt} ] \, .
\eeq
The curly bracket in Eq.~(\ref{cG1}) indicates 
that $\cG_t$ satisfies the equation $\left[ \cG_t, 
\cH_f \right] = (1-f_t^{-1})\cH_t$, which is 
designed according to the similarity renormalization
group procedure described in Ref.~\cite{GlazekWilson}. 
The form factor $f_t$ in Eq.~(\ref{cG1}) is chosen in 
the form that also appears in lowest-order solutions 
obtained using the generator defined in Eq.~(\ref{cG2}).
Namely, in an interaction Hamiltonian term in which 
$R$ and $L$ refer to the effective particles that enter 
and emerge from the interaction, respectively, the 
form factor is
\beq
\label{f}
f_t \es e^{-t(\cM_L^2 - \cM^2_R)^2} \, ,
\eeq
where $\cM_L$ and $\cM_R$ denote the free invariant 
masses of the corresponding particles.

The operator $\cH_f$ in Eq.~(\ref{cG2}) is called 
the free Hamiltonian. It is the part of $\cH_0(a_0)$ 
that does not depend on the coupling constants, 
\beq
\label{cHf} 
\cH_f \es
\sum_i \, p_i^- \, a^\dagger_{0i} a_{0i} \, ,
\eeq 
where $i$ denotes the quantum numbers of gluons 
and $p_i^-$ is the free FF energy for the gluon 
kinematical momentum components $p_i^+$ and 
$p_i^\perp$,
\beq
\label{pi-A}
p^-_i \es { p_i^{\perp \, 2} \over p_i^+} \, .
\eeq
The operator $\cH_{Pt}$ is defined in terms of 
$\cH_t$, the latter considered an arbitrary 
series of normal-ordered powers of the creation 
and annihilation operators,
\beq
\label{Hstructure} 
\cH_t(a_0) =
\sum_{n=2}^\infty \, 
\sum_{i_1, i_2, ..., i_n} \, c_t(i_1,...,i_n) \, \, a^\dagger_{0i_1}
\cdot \cdot \cdot a_{0i_n} \, .
\eeq 
Namely, $\cH_{Pt}$ differs from $\cH_t$ only by 
multiplication of each and every term in it 
by a square of a total $+$ momentum involved
in a term,
\beq
\label{HPstructure} 
\cH_{Pt}(a_0) \es
\sum_{n=2}^\infty \, 
\sum_{i_1, i_2, ..., i_n} \, c_t(i_1,...,i_n) \, 
\left( {1 \over
2}\sum_{k=1}^n p_{i_k}^+ \right)^2 \, \, a^\dagger_{0i_1}
\cdot \cdot \cdot a_{0i_n} \, .
\eeq 
This multiplication implies that $\cH_t$ for all
values of $t$ possesses {\it seven} kinematical
symmetries of the FF Hamiltomian dynamics: three
translations within the front, rotation around
$z$-axis, two transformations generated by $K^1 +
J^2$ and $K^2-J^1$, and the boost generated by 
$K^3$. The latter is the seventh symmetry generator 
that does not have a counterpart in the commonly 
used instant form of dynamics, which has only six 
kinematical symmetries.

Solutions to the RGPEP equation can be found 
using expansion in powers of the coupling 
constant $g$. Such expansion is used in 
Sec.~\ref{ThreeGluonTerm}. The last step in 
the RGPEP is the replacement of bare creation 
and annihilation operators by effective ones. 
One obtains
\beq
\hat H_t \es \cU_t \, \cH_t(a_0) \, \cU_t^\dagger \ .
\eeq
This operator can be used for approximate, 
i.e., neglecting quarks, computations of 
the states of hadrons made of gluons and the 
transition amplitudes for scattering, decay 
and production processes that involve such  
hadrons. The arbitrary parameter $t$ can be 
adjusted in order to reduce the complexity 
of any such calculation to minimum. In practice, 
it means that the size of gluons $s$ is chosen 
to match the inverse of the momentum scale 
that characterizes a process of interest.

Since solutions to the RGPEP equations involve interactions that
are smoothed by form factors, the procedure is thought to provide
a means for the understanding of the connection between
quantum field theory and phenomenological models. For example,
we need to find the mathematical connection between QCD and
the properties of hadrons. This problem needs solution irrespective
of the form of dynamics one uses, e.g. see~\cite{Hilger:2014nma, Hilger:2015hka}.

\section{ Details of the initial Hamiltonian }
\label{initialH}

All interaction terms in the operator $\hat P^-$ of
Eq.~(\ref{hatP-}), are regulated as described in 
Sec.~\ref{reg}. The regularization does not change 
the field operator $\hat A^\mu$ and the associated 
free part $\cH_f$ of the Hamiltonian, besides the 
initial condition that $k^+ > \epsilon^+ \to 0$. 
The latter condition eliminates the terms that 
contain only annihilation or only creation operators. 
If they were present in $\hat P^-$, it would produce 
non-normalizable states by acting on the vacuum 
$|0\rangle$ and all other states built using action 
of creation operators on $|0\rangle$~\cite{DiracQED}. 

The regularized Hamiltonian terms corresponding to 
densities given in Eqs.~(\ref{HA2}) to (\ref{HA4}), 
are listed below in the same order.
\beq 
H_{A^2} \es \sum_{\sigma c} \int [k] {k^{\perp \, 2} \over k^+}
a^\dagger_{k\sigma c}a_{k\sigma c} \ , \\
\label{H3}
H_{A^3} \es \sum_{123}\int[123] \, \tilde \delta(p^\dagger - p)\,\tilde
r_{\Delta\delta}(3,1) \left[g\,Y_{123}\, a^\dagger_1 a^\dagger_2 a_3 +
g\,Y_{123}^*\, a^\dagger_3 a_2 a_1 \right] \ ,  
\eeq
where
\beq
\label{Y123}
Y_{123} = i f^{c_1 c_2 c_3} \left[ \varepsilon_1^*\varepsilon_2^*
\cdot \varepsilon_3\kappa - \varepsilon_1^*\varepsilon_3 \cdot
\varepsilon_2^*\kappa {1\over x_{2/3}} - \varepsilon_2^*\varepsilon_3
\cdot \varepsilon_1^*\kappa {1\over x_{1/3}} \right],  
\eeq
with $\varepsilon \equiv \varepsilon^\perp$ and $ \kappa
\equiv \kappa^\perp_{1/3}$. Symbols $p^\dagger$ and $p$ 
denote the total momenta of created and annihilated particles, 
respectively.
\begin{figure}[h]
 \includegraphics[width=0.3\textwidth]{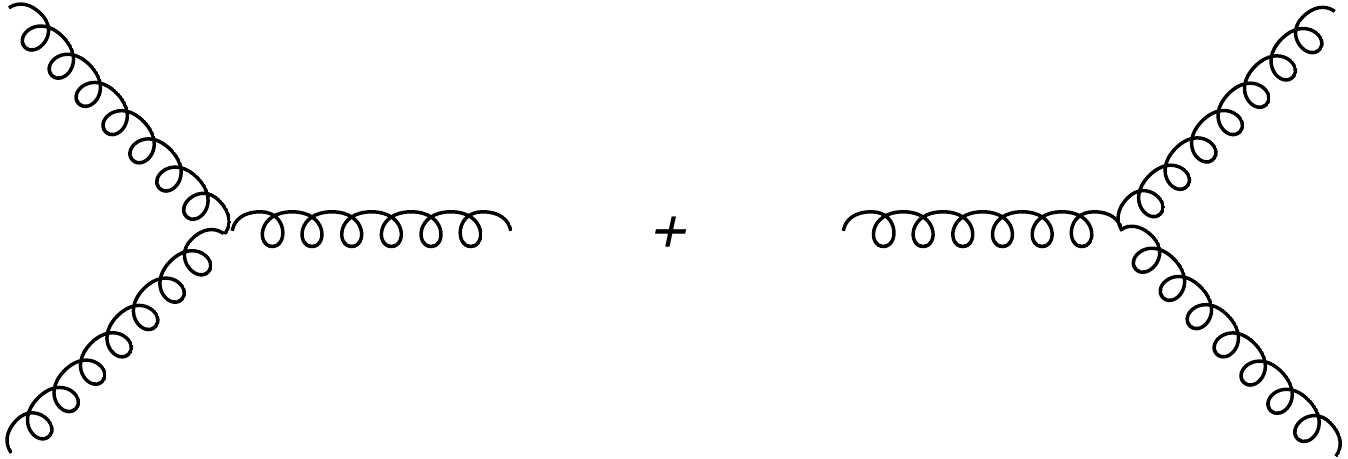}
 \caption{Bare three-gluon vertex.}
\end{figure}
\beq
H_{A^4} = \sum_{1234}\int[1234] \, \tilde \delta(p^\dagger - p)\,{g^2
\over 4}\, \left[ \Xi_{A^4 \, 1234} a^\dagger_1 a^\dagger_2
a^\dagger_3 a_4 + X_{A^4\,1234} a^\dagger_1 a^\dagger_2 a_3 a_4 +
\Xi^*_{A^4\,1234} a^\dagger_4 a_3 a_2 a_1 \right] \, . 
\eeq 
\beq
\Xi_{A^4 \, 1234} \es {2\over 3}
[\tilde r_{1+2,1} \tilde r_{4,3}\,
(\varepsilon_1^*\varepsilon_3^* \cdot \varepsilon_2^*\varepsilon_4 -
\varepsilon_1^*\varepsilon_4 \cdot \varepsilon_2^*\varepsilon_3^*)\,
f^{a c_1 c_2}f^{a c_3 c_4} +
 \tilde r_{1+3,1} \tilde r_{4,2}\,
(\varepsilon_1^*\varepsilon_2^* \cdot \varepsilon_3^*\varepsilon_4 -
\varepsilon_1^*\varepsilon_4 \cdot \varepsilon_2^*\varepsilon_3^*)\,
f^{a c_1 c_3}f^{a c_2 c_4} 
\np
\tilde r_{3+2,3} \tilde r_{4,1}\,
(\varepsilon_1^*\varepsilon_3^* \cdot \varepsilon_2^*\varepsilon_4 -
\varepsilon_3^*\varepsilon_4 \cdot \varepsilon_2^*\varepsilon_1^*)\,
f^{a c_3 c_2}f^{a c_1 c_4}] \, .
\label{Sigma1}
\eeq
\beq
X_{A^4 \, 1234} \es 
\tilde r_{1+2,1} \tilde r_{3+4,3}\,
(\varepsilon_1^*\varepsilon_3 \cdot \varepsilon_2^*\varepsilon_4 -
\varepsilon_1^*\varepsilon_4 \cdot \varepsilon_2^*\varepsilon_3)\,
f^{a c_1 c_2}f^{a c_3 c_4} 
\np
[\tilde r_{3,1} \tilde r_{2,4}+\tilde r_{1,3} \tilde
r_{4,2}]\, (\varepsilon_1^*\varepsilon_2^* \cdot \varepsilon_3
\varepsilon_4 - \varepsilon_1^*\varepsilon_4 \cdot
\varepsilon_2^*\varepsilon_3)\, f^{a c_1 c_3}f^{a c_2 c_4} 
\np
[\tilde r_{3,2} \tilde r_{1,4} + \tilde r_{2,3} \tilde
r_{4,1}] \, (\varepsilon_1^*\varepsilon_2^* \cdot 
\varepsilon_3 \varepsilon_4 - \varepsilon_1^*\varepsilon_3 \cdot
\varepsilon_2^*\varepsilon_4)\, f^{a c_1 c_4}f^{a c_2 c_3} \, .
\label{X1}
\eeq
\beq
H_{[\partial A A]^2} \es
\sum_{1234}\int[1234] \, \tilde \delta(p^\dagger
- p)\,g^2\, \left[ \left( \Xi_{[\partial A A]^2 \, 1234} a^\dagger_1
a^\dagger_2 a^\dagger_3 a_4 + h.c. \right) + X_{[\partial A
A]^2\,1234} a^\dagger_1 a^\dagger_2 a_3 a_4 \right] \,. 
\eeq
\beq
\Xi_{[\partial A A]^2\, 1234} \es
 - {1\over 6}[\tilde r_{1+2,1} \tilde r_{4,3}\,
\varepsilon_1^*\varepsilon_2^* \cdot \varepsilon_3^*\varepsilon_4\,
{(x_1 - x_2)(x_3+x_4)\over (x_1 + x_2)^2}\, f^{a c_1 c_2}f^{a c_3 c_4}
\np
\tilde r_{1+3,1} \tilde r_{4,2}\,
\varepsilon_1^*\varepsilon_3^* \cdot \varepsilon_2^*\varepsilon_4\,
{(x_1 - x_3)(x_2+x_4)\over (x_1 + x_3)^2}\, f^{a c_1 c_3}f^{a c_2 c_4} 
\np
\tilde r_{3+2,3} \tilde r_{4,1}\,
\varepsilon_3^*\varepsilon_2^* \cdot \varepsilon_1^*\varepsilon_4\,
{(x_3 - x_2)(x_1+x_4)\over (x_3 + x_2)^2}\, f^{a c_3 c_2}f^{a c_1
c_4}] \, .
\label{Sigma2}
\eeq
\beq
X_{[\partial A A]^2\, 1234} \es 
{1\over 4}[ \tilde r_{1+2,1} \tilde r_{3+4,3}\,
\varepsilon_1^*\varepsilon_2^* \cdot \varepsilon_3 \varepsilon_4\,
{(x_1 - x_2)(x_3 - x_4)\over (x_1 + x_2)^2}\, f^{a c_1 c_2}f^{a c_3
c_4} 
\nm
[\tilde r_{3,1} \tilde r_{2,4}+\tilde r_{1,3} \tilde
r_{4,2}]\, \varepsilon_1^*\varepsilon_3 \cdot \varepsilon_2^*
\varepsilon_4\, {(x_1 + x_3)(x_2 + x_4)\over (x_2 - x_4)^2}\, f^{a c_1
c_3}f^{a c_2 c_4} 
\nm
[\tilde r_{3,2} \tilde r_{1,4} + \tilde r_{2,3} \tilde
r_{4,1}] \, \varepsilon_1^*\varepsilon_4 \cdot
\varepsilon_2^*\varepsilon_{3}\, {(x_2 + x_3)(x_1+x_4)\over (x_1 -
x_4)^2}\, f^{a c_1 c_4}f^{a c_2 c_3}] \, .
\label{X2}
\eeq
\begin{figure}[h]
 \includegraphics[width=0.5\textwidth]{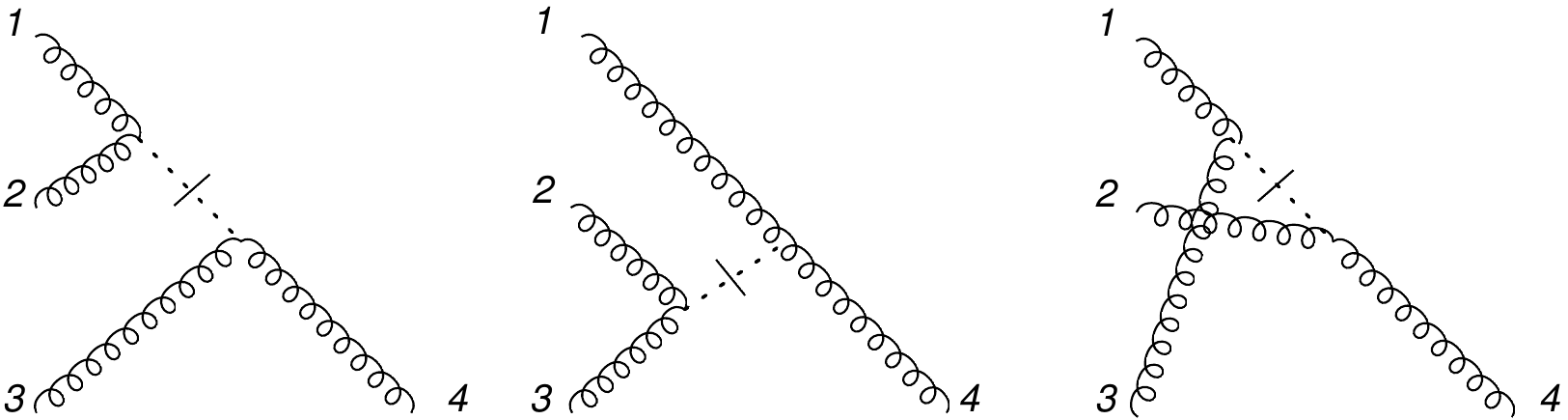}
 \caption{Graphical representation of terms~(\ref{Sigma1}) and~(\ref{Sigma2}).}
 \includegraphics[width=0.4\textwidth]{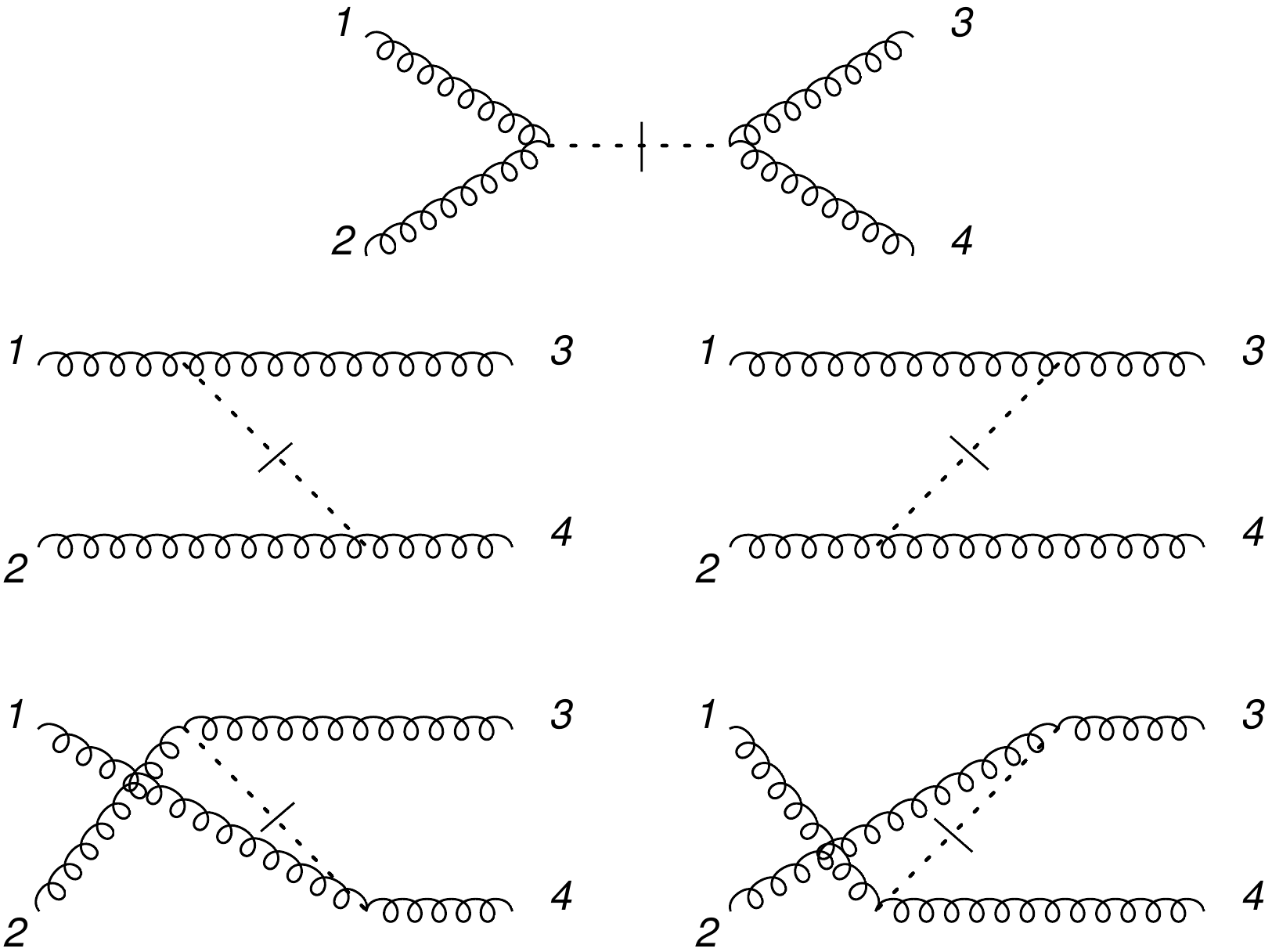}
 \caption{Graphical representation of terms~(\ref{X1}) and~(\ref{X2}).}
\end{figure}
In all these formulas, the dot $\cdot$ is used 
merely to visually separate factors comprised
of scalar products of transverse polarization 
vectors.

\section{Integration of the RG equations order by order}
\label{RGPEPsolution}

The expansion in a series of powers of the coupling 
constant $g$ is inserted in Eq.~(\ref{H'text}) and
solved for the first four terms, i.e., including terms
of order 1, $g$, $g^2$ and $g^3$. We use notation
adopted in Eqs.~(\ref{RGeq1st}) to (\ref{RGeq3rd}).

\subsection{First order terms}
\label{1storderAppendix}

Integration of the terms order $g$ yields
\beq
Y_{21t} + Y_{12t}
\es
f_t \left[ Y_{210} + Y_{120} \right] \ ,
\eeq
where the form factor $f_t$ is given in
Eq.~(\ref{f}). The corresponding Hamiltonian term
is
\beq
H_{(1)} \es \sum_{123}\int[123] \, \tilde \delta(p^\dagger - p) \,
f_t \, {\tilde r}_{\delta}(x_1)
\left[g\,Y_{123}\, a^\dagger_{t 1} a^\dagger_{t 2}
a_{t 3} + g\,Y_{123}^*\, a^\dagger_{t 3} a_{t 2}
a_{t 1} \right] \ .
\eeq
Note the absence of ultra-violet
regularization factors and the 
presence of small-$x$ regularization 
factor ${\tilde r}_{\delta}(x_1)$. 
The reason is that the form factor $f_t$
removes sensitivity to transverse momenta 
much larger than $1/s$ but, for massless 
gluons, does not regulate small-$x$ 
divergences. The creation and annihilation 
operators correspond to the scale parameter $t$.

\begin{figure}[h]
 \includegraphics[width=0.3\textwidth]{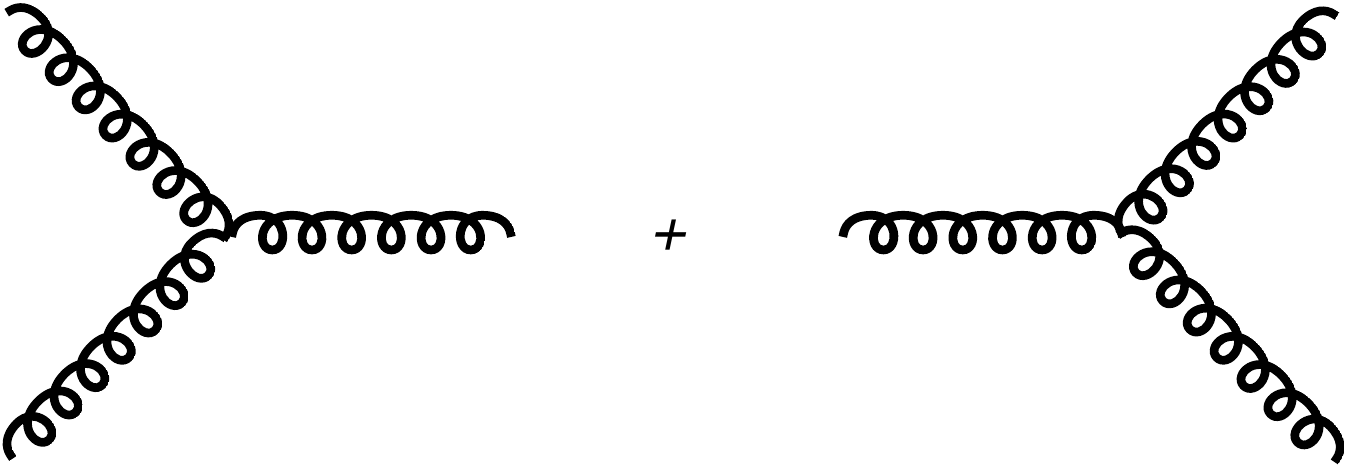}
 \caption{First-order term of the three-gluon vertex.}
\end{figure}

\subsection{Second order terms}
\label{2ndorderAppendix}

\begin{figure}[h]
\includegraphics[width=0.21\textwidth]{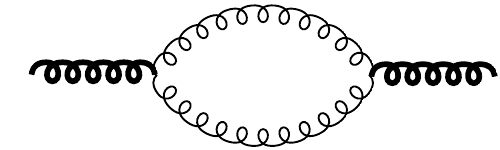}
 \caption{\label{Figgluonmass} Graphical representation 
 of the second-order RGPEP contribution to the 
 Hamiltonian effective gluon mass term. Thin internal 
 lines correspond to intermediate bare gluons and thick 
 external to the effective gluons.}
\end{figure}

Solutions for second-order terms are
\beq
\hat \mu^2_t
\es 
\hat \mu^2_0 +
\int_0^t d\tau
\left[ \left[ E , f_\tau( Y_{21P0} + Y_{12P0})\right], f_\tau(Y_{210} + Y_{120})
  \right]_\mu \ , \\
X_{22t} 
\es
f_t \, \tilde X_0 + 
f_t \int_0^t d\tau
f_\tau^{-1} \left[ \left[ E , f_\tau( Y_{21P0} + Y_{12P0})\right], 
f_\tau(Y_{210} + Y_{120}) \right]_{X_{22}} \ , \\
\Xi_{31t} 
\es
f_t \, \tilde \Xi_{310} + 
f_t \int_0^t d\tau
f_\tau^{-1} \left[ \left[ E , f_\tau Y_{21P0} \right], 
f_\tau Y_{210} \right]_{\Xi_{31}} \ , \\
\Xi_{13t} 
\es
f_t \, \tilde \Xi_{130} + 
f_t \int_0^t d\tau
f_\tau^{-1} \left[ \left[ E , f_\tau Y_{21P0} \right], 
f_\tau Y_{210} \right]_{\Xi_{13}} \ .
\eeq
The gluon-mass term consists of the product of two 
bare vertices, see Fig.~\ref{Figgluonmass},
\beq
\hat \mu^2_t
\es 
\hat \mu^2_0 
-
p^+ 
{1 - f^2_t \over \cM_2^2 } 
\left[   Y_{120} Y_{210} \right]_\mu \ .
\eeq
The subscript $\mu$ indicates that one 
extracts the mass squared term from the 
product of operators in the bracket.
This leads to Eq.~(\ref{gluonmasssol}).

\subsection{Third-order terms}
\label{3rdorder}

The third-order term needed 
for evaluation of the Hamiltonian running coupling is 
\beq
Y_{h21t}
\es
f_t \tilde Y_{h210}
\np
f_t \int_0^t  d\tau \, f^{-1}_\tau 
\left[ \left[ E , X_{22P\tau} \right] f_\tau (Y_{210}) \right]_{ Y_{h21} } 
-
f_t \int_0^t  d\tau \, f^{-1}_\tau 
\left[  f_\tau(Y_{120}) \left[ E , \Xi_{31P\tau} \right] \right]_{ Y_{h21} } 
\np
f_t \int_0^t  d\tau \, f^{-1}_\tau 
\left[ \left[ E , f_\tau(Y_{21P0}) \right], \hat \mu^2_\tau \right]_{ Y_{h21} }
-
f_t \int_0^t  d\tau \, f^{-1}_\tau 
\left[  X_{22\tau} \left[ E , f_\tau(Y_{21P0})\right] \right]_{ Y_{h21} } 
\np
f_t \int_0^t  d\tau \, f^{-1}_\tau 
\left[ \left[ E , f_\tau(Y_{12P0}) \right] \Xi_{31\tau} \right]_{ Y_{h21} } \ .
\label{Yh21t}
\eeq
In an abbreviated notation,
\beq
Y_{h21t} = f_t \sum_n \gamma_{t\,21(n)}
\eeq
where $n$ ranges from $a$ to $i$ and $n=j$ 
for the vertex counterterm. One has
\beq
 \gamma_{t\, 21(n)} \es  \sum_{123}\int[123] \ \tilde \delta(k_1 + k_2 -
k_3)\,\, { g^3 \over 16 \pi^3} \,\, {1 \over 2} \,\, \gamma_{(n)}
\,\,\, a^\dagger_1 a^\dagger_2 a_3 \quad .  
\eeq
We list below results for the vertex functions 
$\gamma_{(n)}$ for all values of $n$ from $a$ to $i$.
The counterterm $\gamma_{(j)}$ is described afterwards
in the next section.

\subsubsection{ Vertex function $\gamma_{(a)}$ }

\beq
\gamma_{(a)} \es   8 \, {N_c \over 2} i f^{c_1 c_2 c_3}
\int_{x_1}^1{dx \,r_{\delta t}(x) \over x(1-x)(x-x_1)} \,
\int d^2\kappa^\perp \,\,r_{\Delta t}(\kappa^\perp)\,\,
{ \cB_{t(a)} \over k^{+ \, 2}_3 }
\,\, \kappa^i_{68} \kappa^j_{16} \kappa^k \,\, \varepsilon^{ijk}_{(a)}
\, + \, (1 \leftrightarrow 2) \ , 
\eeq
where
\beq
r_{\delta t}(x) \es r_\delta(x) \, r_\delta(1-x) \, r_\delta(x_1/x)
\, r_\delta[(x-x_1)/x] \, r_\delta[(x-x_1)/x_2] \, r_\delta[(1-x)/x_2] \ , 
\label{rdt}
\\
 r_{\Delta t}(\kappa^\perp) \es \exp{[-2(\kappa^{\perp \,2}_{68} +
\kappa^{\perp \,2}_{16} + \kappa^{\perp \,2})/\Delta^2]} \ ,   \label{rDt} 
\eeq
\begin{eqnarray}
\varepsilon^{ijk}_{(a)} \es  \varepsilon_1^{*j} \varepsilon_2^{*i}
\varepsilon_3^k
\left[ 1 - {x\over x - x_1} + {1 \over x_1} - { 2 x \over x_1} +
{ x \over (1 - x) x_1} + {x x_2 \over (1 - x) x_1} + {x x_2 \over (x -
x_1) x_1} \right] \nn & & + \varepsilon_1^{*k} \varepsilon_2^{*i}
\varepsilon_3^j
\left[ {1 \over x - x_1} - {1 \over 1 - x} \right] 
+ \varepsilon_1^{*k} \varepsilon_2^{*j} \varepsilon_3^i
\left[ {- x_2\over (1 - x) (x - x_1)} \right]
+ \varepsilon_1^{*i} \varepsilon_2^{*k} \varepsilon_3^j
\left[ {x_2 \over (1 - x) (x - x_1)} \right] \nn
 & & + \varepsilon_1^{*i} \varepsilon_2^{*j} \varepsilon_3^k
\left[{-x_2 \over x - x_1} + {x x_2 \over (1 - x) (x - x_1)} \right]
+ \varepsilon_1^{*j} \varepsilon_2^{*k} \varepsilon_3^i
\left[{-x_2 \over (1 - x) x_1} - {x x_2 \over (1 - x) (x - x_1) x_1} \right]\nn
 & & + \varepsilon_1^* \varepsilon_2^*
\left[\delta^{ik} \varepsilon_3^j {x_2 \over (1 - x)^2} +
\delta^{jk} \varepsilon_3^i {x_2 \over (1 - x) x} +
\delta^{ij} \varepsilon_3^k \left({x x_2 \over (x - x_1)^2} -
 {x_2 \over 1 - x}\right) \right] \nn
 & & + \varepsilon_1^*\varepsilon_3
\left[ \delta^{jk} \varepsilon_2^{*i} \left({x \over (1 - x) (x - x_1)}
- { 1 \over x}\right) -
\delta^{ij} \varepsilon_2^{*k} {x x_2 \over (1 - x) (x - x_1)^2} -
\delta^{ik} \varepsilon_2^{*j} {x x_2 \over (1 - x)^2 (x - x_1)} \right] \nn
 & & +  \varepsilon_2^*\varepsilon_3
\left[\delta^{ij} \varepsilon_1^{*k} {- x_2 \over (x - x_1)^2} +
\delta^{jk} \varepsilon_1^{*i} {x_2 \over x (x - x_1)} -
\delta^{ik} \varepsilon_1^{*j}\left( {x x_2 \over (1 - x)^2 x_1} {+}
{x_2 \over (x - x_1) x_1}\right) \right]
\ ,
\end{eqnarray}
and
\beq
{ \cB_{t(a)} \over k^{+ \, 2}_3 } \es 
- \frac{x \cM_{16}^2-\cM^2}{\cM^4_{16}+\cM^4-\cM^4_{bd}}(x_2\cM^2_{68}+\cM^2_{bd}) 
\left(\frac{f_{16}f f_{68}/f_{12}-1}{\cM^4_{16}+\cM^4+\cM^4_{68}-\cM^4_{12}}-
\frac{f_{68}f_{bd}/f_{12}-1}{\cM^4_{68}+\cM^4_{bd}-\cM^4_{12}}\right) \nn
&& +
\frac{x_2 \cM^2_{68}+x\cM_{16}^2}{\cM_{68}^4+\cM_{16}^4-(\cM^2 - \cM_{12}^2)^2}
(2\cM^2- \cM_{12}^2)
\left( \frac{f f_{68}f_{16}/f_{12}-1}{\cM^4-\cM_{12}^4+\cM_{68}^4+\cM_{16}^4}-
\frac{f_{ca}f/f_{12}-1}{2\cM^2(\cM^2-\cM_{12}^2)}\right) \ . \nn
\eeq

\subsubsection{ Vertex function $\gamma_{(b)}$ }

\begin{equation}
 \gamma_{(b)} =  2 \, {N_c\over 2} i f^{c_1 c_2 c_3}
\int_{x_1}^1{dx\,r_{\delta t}(x) \over x(1-x)} \,
\int d^2\kappa^\perp \,\,r_{\Delta t}(\kappa^\perp) \,\,
{ \cB_{t(b)} \over k^+_3 }\,\,\varepsilon_{(b)} \, + \, (1
 \leftrightarrow 2) \ ,
\end{equation}
where
\beq
\varepsilon_{(b)} \equiv  \varepsilon^\perp_{(b)}\kappa^\perp =
\varepsilon_1^*\varepsilon_2^* \cdot \varepsilon_3\kappa
\left( 1 - s_{(b)} - {1 \over x} - {1 \over 1-x} \right)
+
\varepsilon_1^*\varepsilon_3 \cdot \varepsilon_2^*\kappa
\left({1\over x} + {s_{(b)} \over 1-x}\right)
+
\varepsilon_2^*\varepsilon_3 \cdot \varepsilon_1^*\kappa
\left({1 \over 1 - x } + {s_{(b)} \over x}\right)
\, ,
\eeq
with $ s_{(b)} = (x_1 + x)(x_2 + 1 -x)/(x - x_1)^2$ and
\beq
{ \cB_{t(b)} \over k^+_3 } \es 
 \frac{ 2\cM^2-\cM_{12}^2 }{2\cM^2(\cM^2 - 
\cM_{12}^2)}\left(f_{ca}f/f_{12}-1\right) \ .
\eeq

\subsubsection{ Vertex function $\gamma_{(c)}$ }

\begin{equation}
  \gamma_{(c)} =  2 \, { - N_c\over 2} i f^{c_1 c_2 c_3}
\int_{x_1}^1{dx\,r_{\delta t}(x) \over (x-x_1)(1-x)} \,
\int d^2\kappa^\perp \,\,r_{\Delta t}(\kappa^\perp) \,\,
{ \cB_{t(c)} \over k^+_3 }\,\,\varepsilon_{(c)} \, + \, (1
 \leftrightarrow 2) \ ,
\end{equation}
where
\beq
\varepsilon_{(c)} &\equiv & \varepsilon^\perp_{(c)}\kappa^\perp_{68} \nn
\es
\varepsilon_1^*\varepsilon_2^* \cdot \varepsilon_3\kappa_{68}
\left( {-x_2 \over x-x_1} +{ x_2 s_{(c)} \over 1-x} \right)
 +
\varepsilon_1^*\varepsilon_3 \cdot \varepsilon_2^*\kappa_{68}
\left(- 1 - s_{(c)} + {x_2 \over 1-x} + {x_2 \over x-x_1}\right) \nn
&& +
\varepsilon_2^*\varepsilon_3 \cdot \varepsilon_1^*\kappa_{68}
\left({-x_2 \over 1 - x } + {x_2 s_{(c)} \over x-x_1}\right) \ ,
\eeq
with $ s_{(c)} = (x_1 - x + x_1)(1 -x + 1)/x^2$ and
\beq
{ \cB_{(c)} \over k^+_3 } \es \frac{ x_2  \cM_{86}^2 +  
\cM_{bd}^2}{\cM_{68}^4+\cM_{bd}^4-\cM_{12}^4}\left(f_{68}f_{bd}/f_{12}-1\right) \ .
\eeq 

\subsubsection{ Vertex functions $\gamma_{(d)}$ and $\gamma_{(f)}$ }

\begin{eqnarray*}
 \gamma_{(d)}+ \gamma_{(f)} \es  4 \, N_c Y_{123}
\int_0^1{dx\,r_{\delta \mu}(x) \over x(1-x)} \,
\int d^2\kappa^\perp \,\,r_{\Delta \mu}(\kappa^\perp) \,\,
\kappa^{\perp \, 2}\left[1 + {1\over x^2} + {1\over (1-x)^2}\right]
\left[
{ {\cB}_{t(d)} \over x^2_2 k^{+\, 2}_3 } + { {\cB}_{t(f)} \over  x_2 k^+_3 {\cal M}^2 }
\right] 
\np
4  Y_{123} { {\cB}_{t\,(f)} \over  x_2 k^+_3
}{\tilde \mu}^2_\delta  
+  (1 \leftrightarrow 2) \ ,
\label{gammad}
\end{eqnarray*}
where
\beq
{B_{t(d)} \over k_3^{+2 }  }\es
 -\frac{x_2  \cM^2 -  \cM_{12}^2 }{\cM^4 + \cM_{12}^4 -  \cM_{bd}^4}
\left[\cM^2\left( x_2+\frac{1}{x_2} \right) + \cM_{12}^2 \right]
\left(\frac{f^2-1}{2\cM^4}
-\frac{f f_{bd}/f_{12}-1}{\cM^4 + \cM_{bd}^4 -\cM_{12}^4}\right) \ ,
\eeq 
and 
\beq
{\cB_{t(i)} \over k_3^+ }={ {\cM_{12}^2\over \cM^4}(f^2-1)} \ .
\eeq

\subsubsection{ Vertex functions $\gamma_{(g)}$ and $\gamma_{(i)}$ }

\beq
 & & \gamma_{(g)} + \gamma_{(i)}  =   2 \, N_c Y_{123}
\int_0^1{dx\,r_{\delta \mu}(x) \over x(1-x)} \,
\int d^2\kappa^\perp \,\,r_{\Delta \mu}(\kappa^\perp) \,\,
\kappa^{\perp \, 2}\left[1 + {1\over x^2} + {1\over (1-x)^2}\right]
\left[
{\cB_{t(g)} \over k^{+\, 2}_3 } + { \cB_{t(i)} \over k^+_3 {\cal M}^2 }
\right] 
\np
2 \, Y_{123}\, { \cB_{t(i)} \over k^+_3 }{\tilde
\mu}^2_\delta \, + \, (1 \leftrightarrow 2) \ ,
\eeq
where $r_{\delta \mu}(x)$ is given in Eqs.~(\ref{rdt}), the 
ultra-violet regulator is (\ref{rDt}), $ r_{\Delta \mu}(\kappa^\perp) 
= r^4_\Delta (\kappa^{\perp \, 2})$, the mass counterterm 
contribution stems from $ 2g^2 {\tilde \mu}^2_\delta = 16\pi^3 
\mu^2_\delta$, and
\beq
 { B_{t(g)} \over  k_3^{+2} }\es 
  - \frac{  \cM_{12}^2 +\cM^2 }{ 2\cM^2 \cM_{12}^2}
 (  2 \cM^2 -\cM_{12}^2 ) 
\left[ \frac{f^2 - 1}{2\cM^4}-
\frac{f_{ca}f/f_{12}-1}{2\cM^2(\cM^2-\cM_{12}^2)}\right] \ ,
\eeq
and 
\beq
{\cB_{t(i)} \over k_3^+ }={ {-\cM_{12}^2\over \cM^4}(f^2-1)} \ .
\eeq
Graphs ($e$) and ($h$) result from products of terms $\Xi_{31}$ 
or $X_{22}$ with $Y_{31}$. The former are independent of the 
transverse momentum, and the latter is odd in transverse momentum.
This leads to zero in the integration over $\kappa$.
 \begin{figure}[h]
 \includegraphics[width=0.15\textwidth]{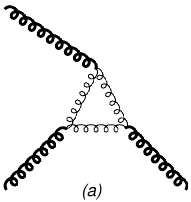}
 \includegraphics[width=0.15\textwidth]{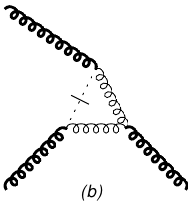}
 \includegraphics[width=0.15\textwidth]{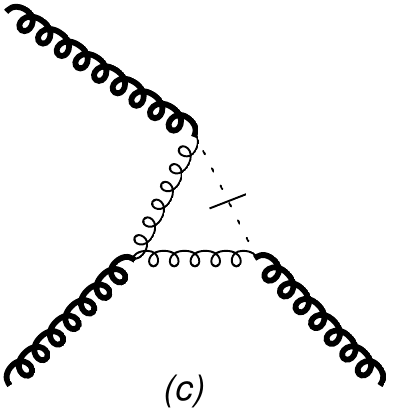}
 \includegraphics[width=0.15\textwidth]{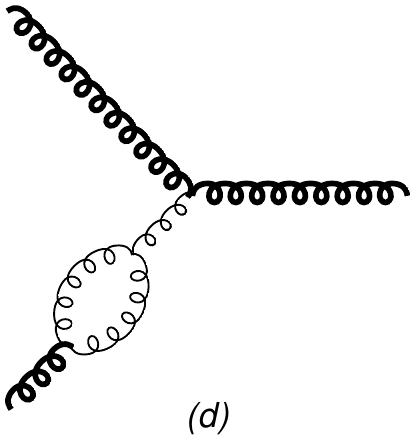}
 \includegraphics[width=0.15\textwidth]{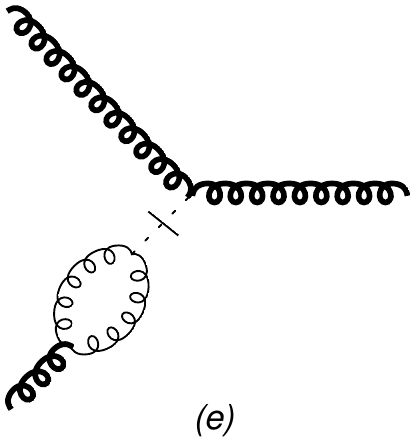}\\
 \includegraphics[width=0.16\textwidth]{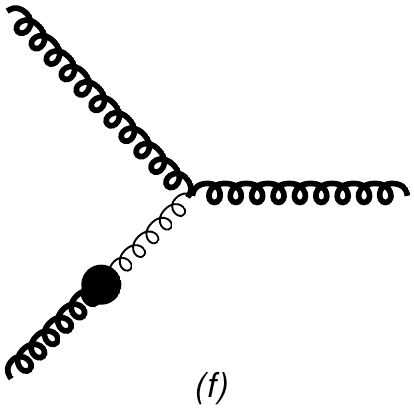}
 \includegraphics[width=0.16\textwidth]{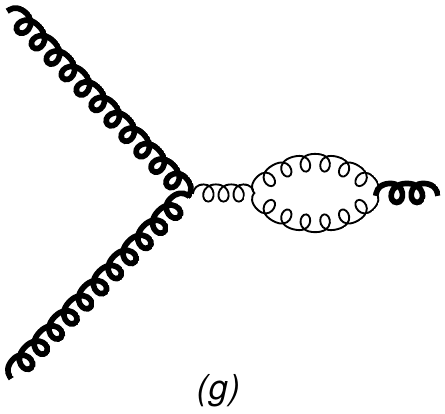}
 \includegraphics[width=0.16\textwidth]{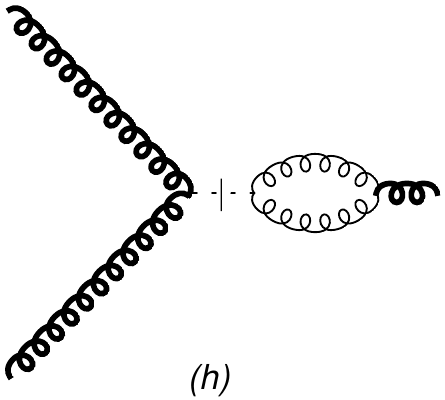}
 \includegraphics[width=0.16\textwidth]{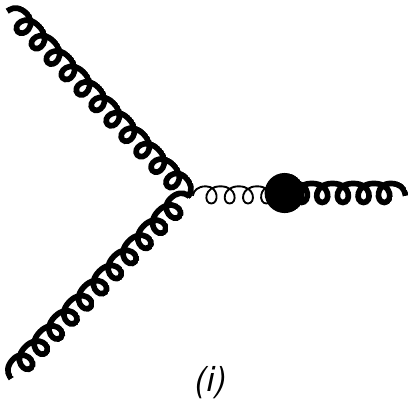}
 \includegraphics[width=0.16\textwidth]{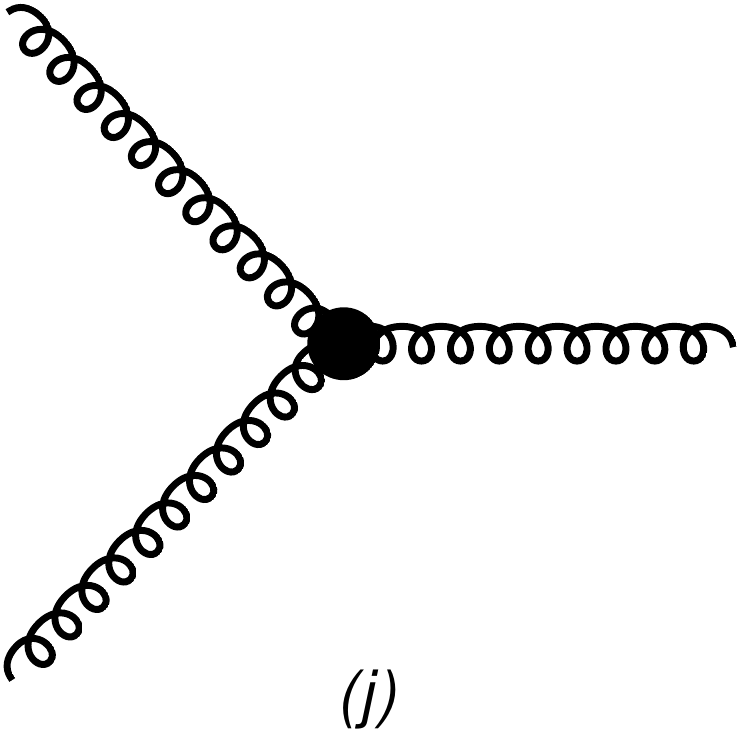}
 \caption{
 \label{Fig3gluonvertex}
 Third-order contributions to the three-gluon vertex, 
 including counterterm, ($j$).}
 \end{figure}

\section{Three-gluon vertex counterterm}
\label{threegCT}

The ultraviolet regularization dependence in 
$\gamma_{(a)}$ turns out to come only from
(see below)
\beq
\left[{ {\cal B}_{t(a)} \over k^{+ \, 2}_3 }\right]_{\Delta}
= {x_2 \over {\cal M}^2 {\cal M}^2_{68}} \ ,  
\eeq
which matches the result found in Ref.~\cite{Glazek:2000dc}. So,
\beq
 \gamma_{(a) \text{div}} =  8 \, {N_c \over 2}i f^{c_1 c_2 c_3}
\int_{x_1}^1 dx \,r_{\delta t}(x) {1-x \over x_2} \,
\left[I^{ijk}\right]_\Delta \varepsilon^{ijk}_{(a)}
\, + \, (1 \leftrightarrow 2) \quad , 
\eeq
where
\beq
i f^{c_1 c_2 c_3}{1-x \over x_2}
\left[I^{ijk}\right]_\Delta \varepsilon^{ijk}_{(a)} =
\pi \ln{\Delta \over |\kappa^\perp_{12}|}  \left[ c_{12}
Y_{12} + c_{13} Y_{13} + c_{23} Y_{23} \right] \quad , 
\eeq
and
\beq
c_{12} \es {2 \over 1-x} + {1 \over x-x_1} + {1 \over x} + {(1-x)^2
\over x_2^2 } - {2 \over x_2} \quad , \label{c12} \\
c_{13} \es {2 \over 1-x} + {1 \over x-x_1} + {1 \over x} + {(1-x)^2
\over x_2 } - 2 \quad , \label{c13} \\
c_{23} \es {2 \over 1-x} + {1 \over x-x_1} + {1 \over x} - {(1-x)^2
\over x_2^2 } - { 1 + x^2 \over x_2} - 2 \quad , \label{c23} 
\eeq
with
\beq
 Y_{12} \es i f^{c_1 c_2 c_3}\varepsilon_1^*\varepsilon_2^* \cdot
\varepsilon_3 \kappa_{12} \quad ,\\
 Y_{13} \es - i f^{c_1 c_2 c_3}\varepsilon_1^*\varepsilon_3 \cdot
\varepsilon_2^* \kappa_{12} {1\over x_{2/3}} \quad , \\
 Y_{23} \es - i f^{c_1 c_2 c_3}\varepsilon_2^*\varepsilon_3 \cdot
\varepsilon_1^* \kappa_{12} {1\over x_{1/3}} \quad .
\eeq
The diverging parts are $\gamma_{(b) \text{div}} =\gamma_{(c) \text{div}} =0$
and $\gamma_{(d)\text{div}} = 2 \gamma_{(g)\text{div}}$, as in 
Ref.~\cite{Glazek:2000dc}, with
\begin{eqnarray}
 & & \gamma_{(g)\text{div}}  =   - N_c Y_{123}
 \tilde r_\delta(x_1)\int_0^1 dx\,r_{\delta \mu}(x)
\int_{\mu^2}^\infty { \pi d\kappa^2 \over \kappa^2 }
\,\,e^{-4\kappa^2/\Delta^2} \,\,
x(1-x)\left[1 + {1\over x^2} + {1\over (1-x)^2}\right] \, + \, (1
 \leftrightarrow 2) \ .
\end{eqnarray} 
These results are the same as in Ref.~\cite{Glazek:2000dc} 
due to the fact that the difference between old and 
new generators resides solely in the RGPEP factors 
$B_t$. The new generator leads to $B_t$s that differ
from old $B_t$s by the additional terms in numerators, 
denominators and arguments of exponentials that depend 
on $\cM_{12}^2$ and do not depend on $\kappa^{\perp}$. 
These additional terms do not affect the behavior 
of $B_t$s when $\kappa^{\perp}\to \infty$. Terms ($i$) 
and ($f$) are not divergent. 

In summary, in the UV-limit of $\kappa^{\perp}\to 
\infty$, the integrands in all vertex functions 
behave in the same way as the corresponding ones 
obtained using the old generator~\cite{Glazek:2000dc}. 
Hence, the divergent part of the vertex counterterm, 
denoted by $\gamma_{\infty \text{div}}$, is also the 
same, c.f. Appendix C in Ref.~\cite{Glazek:2000dc}. 

The divergent part of the vertex counterterm is 
defined by the condition 
\beq
 \gamma_{(a) \text{div}} +3\gamma_{(g)\text{div}} +
\gamma_{\infty \text{div}} + (1 \rightarrow 2)
\es 0 \ .
\eeq
One thus finds the Hamiltonian vertex counterterm 
whose vertex function is 
\beq
\label{gammainfty}
\gamma_\infty = Y_{123} { - N_c \pi \over 3} \, \ln{ \Delta \over
\mu} \left[ 11 + h(x_1) \right]  + \gamma_{\text{finite}} \ ,
\eeq
where $\mu$ denotes the arbitrary separation point 
between the range of integration over large $\kappa^\perp$
that extends up to $\Delta$ and the finite range of integration 
where no dependence on $\Delta$ may arise. The 
finite part of the counterterm, $\gamma_{\text{finite}}$,
removes the artificial dependence on $\mu$.
The function $h(x_1)$ is
\beq
h(x_1) = 6 \int_{x_1}^1 dx \,r_{\delta t}(x) \left[ {2 \over 1-x} +
{1 \over x-x_1} + {1 \over x} \right] - 9 \int_0^1 dx \,r_{\delta
\mu}(x) \left[ {1\over x} + {1 \over 1-x} \right] + (1 \leftrightarrow
2) \ .
\eeq
The vertex function in the complete vertex counterterm is  
\beq
\label{gammajapp}
\gamma_{(j)}=\gamma_\infty  + (1 \rightarrow 2) \ .
\eeq

\section{ Thrid-order contributions to $g_t$ }
\label{Limitk12}

The Hamiltonian coupling constant $g_t$ is
extracted from the term $Y_t$ in Eq.~(\ref{Ytext})
that is linear in $\kappa_{12}^\perp$ in the limit
$\kappa_{12}^\perp\to 0$. Not every term shown in 
Fig.~\ref{Fig3gluonvertex} contributes to the running 
coupling defined this way. Appendix~\ref{3rdorder} 
shows that terms ($e$) and ($h$) do not contribute. 
Furthermore, terms ($b$), ($c$), ($f$) and ($i$) 
vanish faster than linearly in the limit $\kappa_{12}
^\perp\to 0$. Thus, only the terms ($a$), ($g$) and 
($d$) contribute to $g_t$. 
\begin{figure}[h]
 \includegraphics[width=0.8\textwidth]{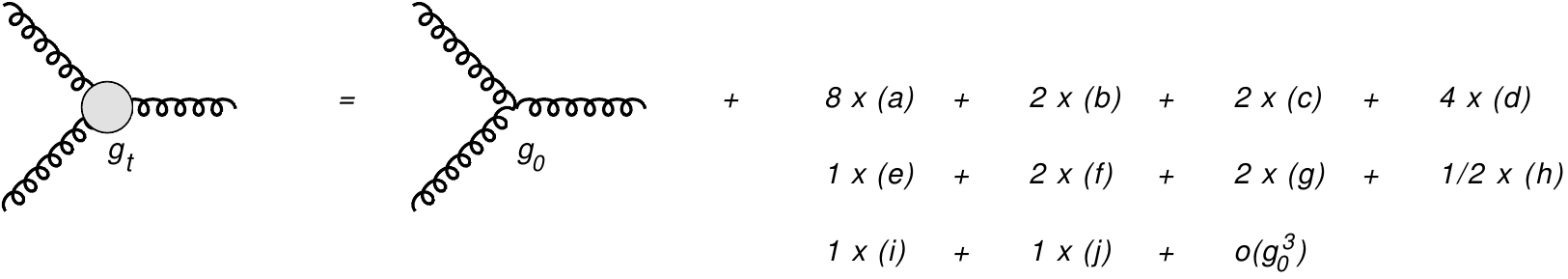}
 \caption{Effective three-gluon vertex (expansion up to third order).}
\end{figure}
Contribution to $g_t$ of each and every term 
is extracted in three steps: 
1) calculation of the coefficient of $\kappa_{12}^\perp$ 
 in the integrand in the limit $\kappa_{12}^\perp \to 0$;
2) integration over $\kappa^\perp$;
3) integration over $x$.
Contributions of the counterterm are defined by 
the subtraction at $t=t_0$ that is described in 
Secs.~\ref{gvct} and~\ref{runningcoupling}.

\subsection{ Contribution to the running coupling from $\gamma_{(a)}$ }
\label{Aa}

The expansion in small $\kappa_{12}^i$ in term ($a$)
concerns the factor
\begin{equation}
{ \cB_{t(a)} -\cB_{t_0(a)}\over k_3^{+2} } \, \kappa^i_{68} \kappa^j_{16} \kappa^k \ ,
\end{equation}
which leads to
\begin{equation}
 { \cB_{t(a)} -\cB_{t_0(a)}\over k_3^{+2} } |_{\kappa_{12}^{\perp}=0} \ 
\left( \kappa^i    \kappa^k \kappa_{12}^j + \frac{(1-x)}{x_2}  \frac{x_1}{x} \kappa^j  \kappa^k \kappa_{12}^i
 \right) 
 \ - \ 
\frac{x_1}{x} \ { \cB'_{t(a)} -\cB'_{t_0(a)}\over k_3^{+2} }|_{\kappa^{\perp}_{12}=0} \kappa_{12}^\perp \  \kappa^i\kappa^j   \kappa^k
 \end{equation}
This is integrated over $\kappa^\perp$, yielding
\beq
\int d^2 \kappa^\perp 
\lim_{\kappa_{12}^i\to 0} 
\left[  { \cB_{t(a)} -\cB_{t_0(a)}\over k_3^{+2} } \, \kappa^i_{68} \kappa^j_{16} \kappa^k \right]
\es 
\cA_1(x,x_1)\, \pi\left( \delta^{ik} \kappa_{12}^j + \frac{(1-x)}{x_2}  \frac{x_1}{x} \delta^{jk}  \kappa_{12}^i
 \right) 
 \frac{1}{4} \ln  \frac{t}{t_0}
 \nm   \frac{x_1}{x}\cA_2(x,x_1) {1 \over x-x_1}\frac{\pi}{4} \left( \delta^{jk} \kappa^{i}_{12}
+\delta^{ik}  \kappa^{k}_{12}+\delta^{ij} \kappa^{k}_{12}
\right)
\, \frac{1}{4} \ln  \frac{t}{t_0} \ ,
\eeq
where $\cA_1(x,x_1) = x^ 2 (1-x)^2{x-x_1 \over x x_2} $ 
and $\cA_2(x,x_1)= 2x^3 (1-x)^3\,{(x-x_1)^2 \over x_2^2 x^2}$. 
Here we have made use of
the formula in Appendix~\ref{Formulas}. The running coupling contribution is extracted from 
\beq
\lim_{\kappa_{12}^i\to 0} { g^3 \over 16 \pi^3} \,\, {1 \over 2} \,\, (\gamma_{t, a} -\gamma_{t_0, a}) \rs  
{ g^3 \over 16 \pi^3} \,\, {1 \over 2} \,\,8 \, {N_c \over 2} i f^{c_1 c_2 c_3} 
\ \lim_{\kappa_{12}^i\to 0} \ c_{t, a}  (x_0, \kappa_{12}^\perp) \ ,
\eeq
where
\beq
\lim_{\kappa_{12}^i\to 0} c_{t, a} (x_1, \kappa_{12}^\perp) 
\es
 \frac{\pi}{4} \ln  \frac{t}{t_0} \int_{x_1}^1 dx \,r_{\delta t}(x)
  \, { (1-x)\over  x_2}\left[
 \left(1- { x_1\, (1-x) \over x_2 x}  \frac{1}{2} \right)
   \varepsilon^{iji}_{(a)}\, \kappa_{12}^j  
 + {1\over 2}{x_1  (1-x)\over x x_2} 
  \kappa_{12}^i  \varepsilon^{ijj}_{(a)}
\right.
 \nm \left.
 { x_1\,(1-x)\over x_2 x}  \frac{1}{2}    
 \varepsilon^{iik}_{(a)}\, \kappa^{k}_{12}
 \right] \,\, 
\, + \, (1 \leftrightarrow 2) + o(\kappa_{12}^\perp) \ .
\eeq
The contraction of indices of the tensor structure $\varepsilon_{(a)}^{ijk}$ 
simplifies the limit of $c_{t, a} (x_1, \kappa_{12}^\perp)$ to
\beq
c_{t, a} (x_1, \kappa_{12}^\perp) \to 
{\pi \over 4}\ln{t\over t_0}
\left[
\int_{x_1}^1 dx\  r_{\delta Y}\ f_{12}^\perp(x, x_1,\varepsilon^\perp) 
\ + \ ( 1 \leftrightarrow 2) 
\right] \ \kappa_{12}^\perp
\ ,
\eeq
where
\beq
f_{12}^\perp(x, x_1,\varepsilon^\perp)
\es
c_{12} \ \varepsilon_1^* \varepsilon_2^* \ \varepsilon_3^\perp
-
c_{13} \ \varepsilon_1^* \varepsilon_3   \ \varepsilon_2^{*\perp} / x_2
-
c_{23} \ \varepsilon_2^* \varepsilon_3   \ \varepsilon_1^{*\perp} / x_1 \ ,
\eeq
with the coefficients $c_{ij}$ given in Eqs.~(\ref{c12})-(\ref{c23}).
Finally, the integration over $x$ leads to the running 
coupling contribution of term ($a$),
\beq
g_{t,(a)} \es 
g_{t_0,(a)} + { g^3 \over 48 \pi^2 } { 1\over 2 }  N_c \, 
\left[ - 11 + 3 \chi_a (x_0) \right]\, \ln{t\over t_0} \ ,
\eeq
with 
\beq
\chi_a(x_0) \es 
\int_{x_0}^1 dx \  r_{\delta Y} \left[ 2/(1-x) + 1/(x-x_0) + 1/x \right] + ( x_0 \rightarrow 1-x_0) \ .
\eeq

\subsection{ Contribution to the running coupling from $\gamma_{(d)}$ }
\label{Ad}

In this case $\cM_{68}^2=\cM^2$, $\cM_{bd}^2=\cM^2/x_2+\cM_{12}^2$. 
The limit $\kappa_{12}^\perp\to 0$ in Eq.~(\ref{gammad}) concerns the factor,
\beq
\lim_{\kappa_{12}^\perp\to0}{\cB_{t\,(d)} - \cB_{t_0\,(d)} \over x_2^2 k_3^{+2 }  } 
& = &
\frac{  1  }{1  -  1/x_2^2}
\left( 1+\frac{1}{x_2^2} \right)  
x^2(1-x)^2
\left(\frac{f_{0}^2- f^2}{2\kappa^4}
-\frac{f_{0}f_{bd,0}-ff_{bd}}{\kappa^4(1 + 1/x_2^2) }\right) \ . 
\eeq
Integration over $\kappa^\perp$ yields:
\beq
\int d^2\kappa^\perp \ \kappa^{\perp 2} 
\lim_{\kappa_{12}^\perp\to0}
\left[{\cB_{t\,(d)} - \cB_{t_0\,(d)} \over x_2^2 k_3^{+2 }  } \right]
\es
-  \   x^2(1-x)^2 {\pi \over 4}\ln{t\over t_0} \ .
\eeq
The last step is the integration over $x$ of 
$\gamma_{(d)} - \gamma_{0,\,(d)}$ in this limit.
The corresponding contribution to the running coupling is
\beq
g_{t, \, (d)} \es 
g_{t_0, \, (d)} 
 - { g^3 \over  16 \pi^2 } \, N_c\, \ln{t\over t_0}
 \left\{ - { 11 \over 6 } +\int_0^1dx\,r_{\delta \mu}(x)  \, \left[ {1 \over x} + { 1 \over 1 - x}\right] \right\} \ .
\eeq

\subsection{ Contribution to the running coupling from $\gamma_{(g)}$ }
\label{Ag}

The calculation is analogous to the previous cases. 
The limit  $\kappa_{12}^\perp\to 0$ concerns the 
difference of renormalization group factors and produces
\beq
\lim_{ \kappa_{12}^\perp\to 0}{ B_{t\,(g)} - B_{t_0\,(g)} \over  k_3^{+2} }
& = &
 - \ t_0\, f_0^2 + \, t\, f^2 - {f_0^2 - f^2\over 2\cM^4} \ .
\eeq
Integration over $\kappa^\perp$ of the first two terms gives zero, 
and the only contributing part is
\beq
\int d^2\kappa^\perp \, \kappa^{\perp \, 2} 
\lim_{ \kappa_{12}^\perp\to 0} \left[{ B_{t\,(g)} - B_{t_0\,(g)} \over  k_3^{+2} }\right]
\es - { x^2(1-x)^2 }  { \pi \over 4} \ln { t \over t_0} \ . 
\eeq
The resulting contribution to the running coupling of term ($g$) is
\beq
 g_{t, \, (g)} \es  
 g_{t_0, \, (g)} 
- { g^3 \over  16 \pi^2 }
  \, N_c \, { 1 \over 2} \ln { t \over t_0}
\left\{ - { 11 \over 6 } +\int_0^1dx\,r_{\delta \mu}(x)  \, \left[ {1 \over x} + { 1 \over 1 - x}\right] \right\} \ .
\eeq

\subsection{ Sum of contributions in App.~\ref{Aa}, \ref{Ad} and \ref{Ag} }
\label{Asum}

Denoting $g_{t_0}$ by $g_0$, the sum of contributions 
($a$), ($d$) and ($g$) up to order $g_0^3$ gives
\beq
g_t \es
g_0 - { g_0^3 \over 48 \pi^2 }   N_c \,  \left[ 11 + h(x_0) \right]\,\ln { \lambda \over \lambda_0} \ .
\eeq 

\subsection{Useful formula}
\label{Formulas}

Integrals of differences of exponentials
that appear in the RGPEP for massless
quanta, are evaluated taking advantage 
of the formula
\begin{equation}
\label{Formula}
 \int d^2 \kappa^{\perp} {f - f_0 \over
\kappa^{\perp 2}} = \frac{\pi}{2} \ln \frac{t_0}{t} \ .
\end{equation}
This formula is a consequence of the RGPEP design
that secures absence of large perturbative
contributions in the matrix elements near diagonal
of the effective Hamiltonian matrix evaluated in
the basis of the Fock space built using creation
operators for effective particles~\cite{Glazek:2012qj}.
Namely, the arguments of form factors $f$ vanish
quadratically as functions of the corresponding 
perturbative denominators.

\end{appendix}


\end{document}